\documentclass[twocolumn,twocolappendix]{aastex631}
\usepackage{comment}
\usepackage{chemformula}
\usepackage{booktabs}
\usepackage{rotating}
\usepackage{mathtools}
\usepackage{graphicx}
\usepackage{soul}
\usepackage{hyperref}
\usepackage{natbib}

\shorttitle{NH$_3$ Formation on Water Ice}
\shortauthors{Ferrero et al.}
\graphicspath{{./}{Figures/}}

\begin{document}

\title{Where does the energy go during the interstellar NH$_3$ formation on water ice? A computational study}

\correspondingauthor{Albert Rimola, Mariona Sodupe}

\author[0000-0001-7819-7657]{Stefano Ferrero}
\affiliation{Departament de Qu\'{i}mica, Universitat Aut\`{o}noma de Barcelona, Bellaterra, 08193, Catalonia, Spain}

\author[0000-0002-2457-1065]{Stefano Pantaleone}
\affiliation{Dipartimento di Chimica and Nanostructured Interfaces and Surfaces (NIS) Centre, Universit\`{a} degli Studi di Torino, via P. Giuria 7, 10125, Torino, Italy}

\author[0000-0001-9664-6292]{Cecilia Ceccarelli}
\affiliation{Univ. Grenoble Alpes, CNRS, Institut de Plan\'{e}tologie et d'Astrophysique de Grenoble (IPAG), 38000 Grenoble, France}

\author[0000-0001-8886-9832]{Piero Ugliengo}
\affiliation{Dipartimento di Chimica and Nanostructured Interfaces and Surfaces (NIS) Centre, Universit\`{a} degli Studi di Torino, via P. Giuria 7, 10125, Torino, Italy}

\author[0000-0003-0276-0524]{Mariona Sodupe}
\affiliation{Departament de Qu\'{i}mica, Universitat Aut\`{o}noma de Barcelona, Bellaterra, 08193, Catalonia, Spain}
\email{mariona.sodupe@uab.cat}

\author[0000-0002-9637-4554]{Albert Rimola}
\affiliation{Departament de Qu\'{i}mica, Universitat Aut\`{o}noma de Barcelona, Bellaterra, 08193, Catalonia, Spain}
\email{albert.rimola@uab.cat}



\begin{abstract}

In the coldest (10--20 K) regions of the interstellar medium, the icy surfaces of interstellar grains serve as solid-state supports for chemical reactions. Among their plausible roles, that of third body is advocated, in which the reaction energies of surface reactions dissipate throughout the grain, stabilizing the product. This energy dissipation process is poorly understood at the atomic scale, although it can have a high impact on Astrochemistry. Here, we study, by means of quantum mechanical simulations, the formation of NH3 via successive H-additions to atomic N on water ice surfaces, paying special attention to the third body role. We first characterize the hydrogenation reactions and the possible competitive processes (i.e., H-abstractions), in which the H-additions are more favourable than the H-abstractions. Subsequently, we study the fate of the hydrogenation reaction energies by means of ab initio molecular dynamics simulations. Results show that around 58--90\% of the released energy is quickly absorbed by the ice surface, inducing a temporary increase of the ice temperature. Different energy dissipation mechanisms are distinguished. One mechanism, more general, is based on the coupling of the highly excited vibrational modes of the newly formed species and the libration modes of the icy water molecules. A second mechanism, exclusive during the NH$_3$ formation, is based on the formation of a transient H$_3$O$^+$/NH$_2^-$ ion pair, which significantly accelerates the energy transfer to the surface. Finally, the astrophysical implications of our findings relative to the interstellar synthesis of NH$_3$ and its chemical desorption into the gas are discussed.

\end{abstract}



\section{Introduction} \label{sec:intro}

Dust grains are the primarily solid-state particles of the interstellar medium (ISM). Their chemical composition, as can be inferred by observationally  IR spectroscopic measurements, is based on a core of silicates or carbonaceous materials \citep{jones2013evolution, henning2010cosmic, jones2017global}, coated by ice mantles predominantly made by water. These ice mantles are usually referred to as “dirty” ices because, beside water, they also contain several other volatile molecules, e.g., CO, CO$_2$, CH$_4$, NH$_3$, CH$_3$OH, among others \citep{boogert2015observations}. Spectroscopic comparisons between spectra from laboratory ice analogs and observational spectroscopic measurements suggest that interstellar ices have an amorphous structure and are relatively porous \citep{watanabe2008ice, fraser2004using, potapov2020ice}. It has been long recognized that interstellar dust grains are of extreme importance for the chemistry of the ISM as, among different roles (e.g., chemical catalysts), they can act as third bodies (or energy sinks), in which the energy released by the reactions occurring on the grain surfaces is dissipated throughout the grains, this way stabilizing the newly formed products. The third body effect of the interstellar grains on chemical reactions is a subject of interest in Astrochemistry because it has direct implications for the occurrence of chemical desorption (CD) of species synthesized on the grain surfaces. CD is a desorption mechanism driven by chemistry, in which an adsorbed molecule on the grain surface is ejected into the gas phase because, once formed on the surface, it desorbs due to the local heating caused by the exothermicity of the reaction. Since the interstellar temperatures are very cold (usually below the sublimation of the icy species, making standard thermal desorption inoperative), CD is an important and frequently resorted mechanism accounted for in the astrochemical models to justify the abundances of gas-phase species that otherwise would not be possible without being considered.

Hydrogenation reactions are the most common interstellar reactions taking place on the surfaces of dust grains \citep{watanabe2008ice, hama2013surface}, because atomic hydrogen is the most abundant element in space and the most mobile atom that can diffuse over the grain surfaces at the ultracold temperatures of the ISM (10--20 K). An emblematic grain surface reaction is the formation of H$_2$, the most abundant molecule in the Universe, which is formed by the coupling of two H atoms \citep{wakelam2017h2, vidali2013h2}. In the gas phase, the coupling of two H atoms in their electronic ground states does not take place because, in the absence of a third body (e.g., a grain), the reaction energy associated with the H-H chemical bond formation cannot be released, consequently leading to its re-dissociation. 
The partitioning and dissipation of reaction energies are key surface phenomena in the synthesis of compounds taking place in the presence of interstellar grains, such as the formation of simple but relevant molecules (e.g., H$_2$, H$_2$O, CH$_3$OH) \citep{wakelam2017h2, vidali2013h2, van2013interstellar, dulieu2010experimental, watanabe2004hydrogenation, fuchs2009hydrogenation, qasim2018formation, watanabe2002efficient, rimola2014combined} and of the so-called interstellar complex organic molecules (iCOMs) \citep{ceccarelli2017seeds, zamirri2019quantum, gutierrez2021icom,perrero2022non, enrique2019reactivity, enrique2022quantum}, these later ones inferring a primogenital organic chemistry to the ISM \citep{ceccarelli2022organic}. However, energy partitioning and dissipation are hitherto poorly understood at an atomistic level. Nevertheless, ab initio molecular dynamics (AIMD) based on Density Functional Theory (DFT) are very powerful computational simulations that allow us to gain insights into these processes because they are based on electronic structure calculations (and hence can deal with the breaking/formation of chemical bonds) and naturally account for anharmonic effects, including the coupling amongst the vibrational modes of the molecule/surface system. For large atomistic models, this methodology becomes computationally very expensive and therefore true statistics of the processes under study cannot be provided. This can only be possible by means of classical molecular dynamics (MD) simulations, in which hundreds of simulations with long simulation timescales can be executed, this way allowing a statistical treatment of the results \citep{fredon2017energy, fredon2018molecular}. However, simulations based on classical force fields cannot cope with the breaking/formation of chemical bonds, and accordingly they are not suitable to investigate the dissipation of energies arising from chemical reactions of molecules synthesis. To cope with these problems, in addition to the AIMD simulations \citep{pantaleone2020chemical, pantaleone2021h2}, other sophisticated approaches have been applied to the study of energy dissipation at surfaces, such as reactive classical MD \citep{pezzella20192, upadhyay2021energy}, embedding schemes \citep{meyer2014modeling, rittmeyer2018energy}, non-equilibrium MD \citep{melani2021vibrational}, and also MD based on neural network potentials \citep{shakouri2018analysis}.

In the present work, we have studied the formation of interstellar NH$_3$, one of the components forming part of the ice mantles, from the successive hydrogenation (i.e., H-addition) reactions of atomic N on water ice mantles, with the particular focus to analyze, at a molecular scale, how the reaction energies of each elementary step partition and dissipate and identify the mechanisms through which these events take place, all in all by means of AIMD simulations.

\section{Methodology} \label{sec:methodology}
\subsection{Computational details}
A set of static calculations, both in the gas phase and on the solid ice surfaces (crystalline and amorphous water ices, see below for more details), were performed with the CP2K software package, using the Quickstep module \citep{kuhne2020cp2k}. We used the DFT PBE (Perdew-Burke-Ernzerhof) method \citep{perdew1996generalized}, plus the a posteriori Grimme’s D3(BJ) correction for dispersion interactions \citep{grimme2010consistent, grimme2011effect}. Core electrons were described by the Goedecker-Teter-Hutter pseudopotentials \citep{goedecker1996separable}, whereas valence electrons with a triple zeta MOLOPT basis set \citep{vandevondele2007gaussian} including a single polarization function (TZVP) with mixed Gaussian Plane Waves (GPW) approach \citep{lippert1997hybrid}. The plane-wave cut-off was set to 600 Ry. In geometry optimizations, the reactive species were optimized on the thermalized slab, in which the atoms of the water molecules of the ice were fixed to keep the 10K-thermalized structure.

The Orca software \citep{neese2020orca} was also employed to check the accuracy of PBE-D3(BJ) for the energetics of the reactions in the gas-phase. Calculations using the range separated hybrid functional $\omega$B97x-D3(BJ) \citep{najibi2018nonlocal}, employing a def2-TZVP basis set \citep{weigend2005balanced} have been carried out, as well as single point energy evaluations at CCSD(T)-F12 level of theory \citep{hattig2012explicitly}, employing a cc-pVTZ-F12 basis set with an auxiliary cc-pVTZ-F12-CABS basis set \citep{Peterson_CCSDT_basis, valeev2004improving}, to check the reliability of both DFT methods. All the calculations have been performed within the unrestricted formalism using the broken symmetry approach for biradical systems \citep{neese2004definition}. Results indicate that both methods provide similar results to CCSD(T) for reaction energies, hence demonstrating the robustness of the PBE-D3(BJ) method for the subsequent energy dissipation study with AIMD simulations at this theory level. 

AIMD simulations were employed to study the fate of the energy released along the various reactions. For these simulations, we only focused on the amorphous water ice surface model, which was previously thermalized at 10K in the NVT ensemble for 1ps (using a CSVR thermostat \citep{bussi2007canonical} to ensure a sampling at a constant T) with a time step of 1fs. The energy dissipation process was studied by means of NVE AIMD simulations following the energy released by every reaction and how this energy is partitioned between the newly formed molecules and the surface, in a similar way to the works by some of us \citep{pantaleone2020chemical, pantaleone2021h2}. To estimate observables, moving averages of 50 MD steps have been calculated. The starting velocities of the water molecules belonging to the ice surface were recovered from those of the NVT thermalization procedure. In the NVE simulations, the time step was lowered to 0.5 fs to ensure a correct energy conservation of the system. 

Finally, in order to gain insights into the energy dissipation mechanisms, we calculated the vibrational density of states (VDOS) for the water molecules of our surface model, as the Fourier transform of the velocity autocorrelation function (equation 1) of our trajectories using the TRAVIS analyser \citep{brehm2011travis, brehm2020travis}.

\begin{equation}
P(\omega, t) = \int_{t_1}^{t_2} \langle \textbf{v}(t) \cdot \textbf{v}(t+ \tau)  \rangle e^{- i \omega t} d\tau
\end{equation}
The VDOS is a powerful tool to assess the energy relaxation because it allows highlighting the surface vibrational modes that are excited by the dissipation of energy from the newly formed molecule. Therefore, it has been here employed to qualitatively understand the vibrational couplings amongst the components of systems.  

\subsection{Water ice surface models}
\begin{figure}
    \centering
    \includegraphics[width=\linewidth]{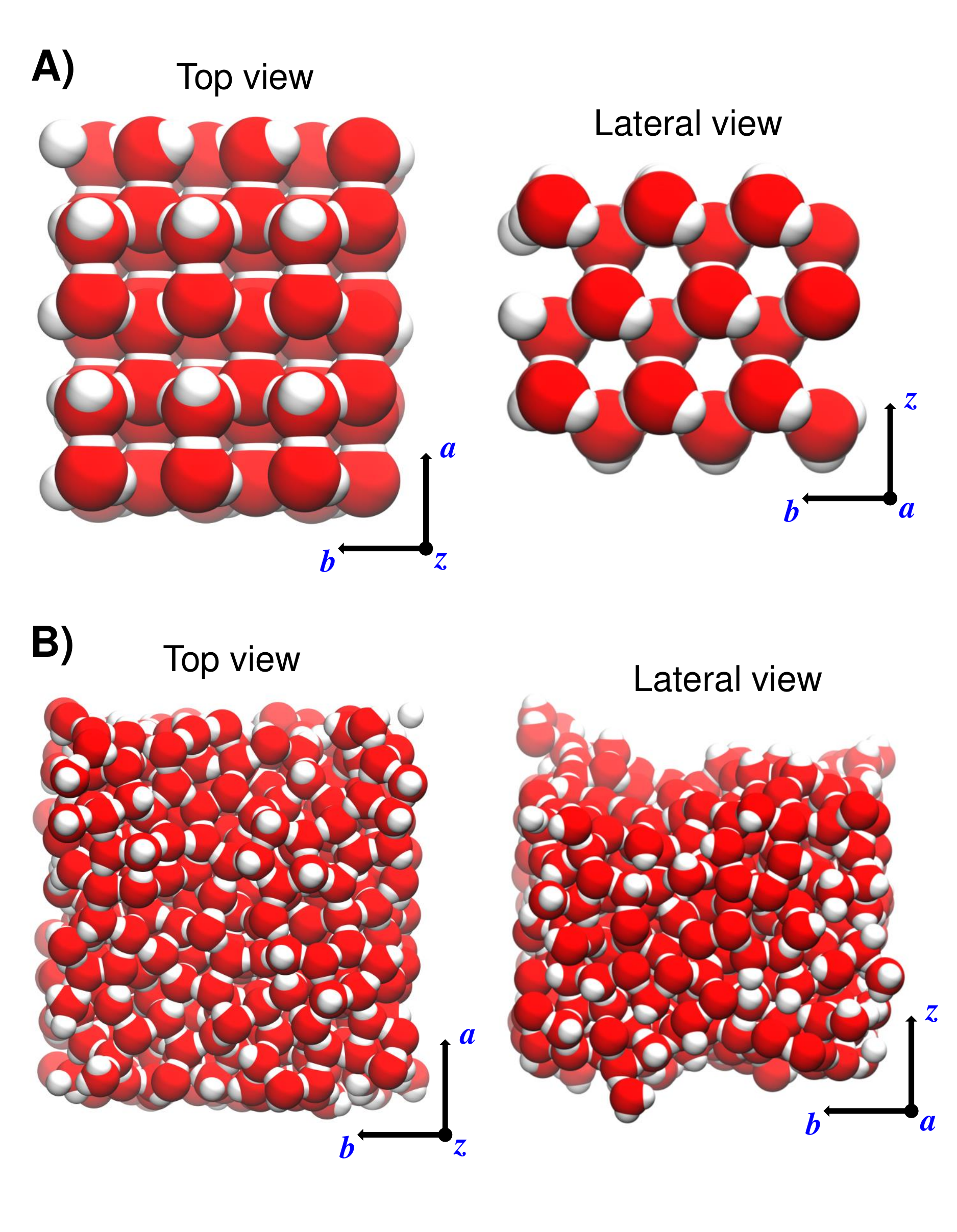}
    \caption{Water ice surface models used in this work. A) crystalline model of the (100) water ice cut out from the proton ordered P-ice bulk. B) amorphous surface model derived from the amorphization of the crystalline surface (see text for more details).}
    \label{fig:1}
\end{figure}

In this work, two different models were used to mimic interstellar water ice surfaces. The first (and simplest) model is a crystalline slab constructed by cutting out the proton ordered P-ice bulk along the (100) plane \citep{casassa1997proton}. This periodic ice surface model has 72 water molecules in the unit cell (216 atoms) and the cell parameters are a=13.156 Å, b=14.162 Å, and $\alpha$=$\beta$=$\gamma$= 90 degrees. The c value, which modulates the interlayer distance among replicated slab images, was fixed to 30 Å, thus ensuring an empty space (about 22 Å) among fictitious slab replicas. This first model was mainly used for static calculations characterizing the potential energy surfaces of the H-addition and H-abstraction reactions.  Figure \ref{fig:1}A shows the structure of the crystalline ice surface slab model.

To study the dissipation and partition of the nascent reaction energies with (NVE) AIMD simulations, we used a larger and amorphous slab model. Larger with the purpose to avoid non-physical temperature increases caused by size limitations of the surface ice model, and amorphous with the purpose to simulate as much realistic as possible interstellar water icy mantle. To have a robust ice surface model accounting for the first aspect, we estimated the temperature increase of the entire system caused by the occurrence of the chemical reaction through the equipartition theorem, in which such an increase of temperature ($\Delta T$) is given by

\begin{equation}
    \Delta T = \frac{2 E_{nasc}}{RN_{at}}
\end{equation}
where $E_{nasc}$ is the nascent energy due to the reaction leading to the chemical bond formation (in this work a N-H for each hydrogenation step) and $N_{at}$ is the number of atoms of the system. According to our calculations, the reaction energies lay between -340 and -440 kJ mol$^{-1}$. These values are similar to the reaction energy of the H$_2$ formation through the coupling of two H atoms. Accordingly, the expected $\Delta T$ is the same (30K at most) and thus, we used the same amorphous water ice surface model adopted by some of us to investigate the fate of the reaction energy in the interstellar H$_2$ formation \citep{pantaleone2021h2}. This slab model was initially built up as a 2x2 supercell of the crystalline slab (a=26.32 Å, b=28.33 Å, and $\alpha$=$\beta$=$\gamma$= 90 degrees), in which more layers of water molecules along the z axis (the non-periodic direction) were added, resulting in 576 water molecules in the unit cell (1728 atoms). Because of these additions, the c value was enlarged to 50 Å (about 30 Å of real vacuum space). Such a crystalline supercell was then “amorphized” by a classical MD annealing process using the rigid TIP3P potential \citep{jorgensen1983comparison}. That is, a first run of 200ps was performed in the NVT ensemble at 300K to break the crystallinity of the initial structure, followed by a second NVT run at 10K for 200 ps to cool down the ice at this interstellar temperature. The final disordered ice structure was optimized at the PBE-D3(BJ) level and then thermalized as described above. Figure \ref{fig:1}B shows the structure of the amorphous ice surface slab model.

\section{Results} \label{sec:results}
\subsection{Energetics of the reactions}

\begin{figure}
    \centering
    \includegraphics[width=\linewidth]{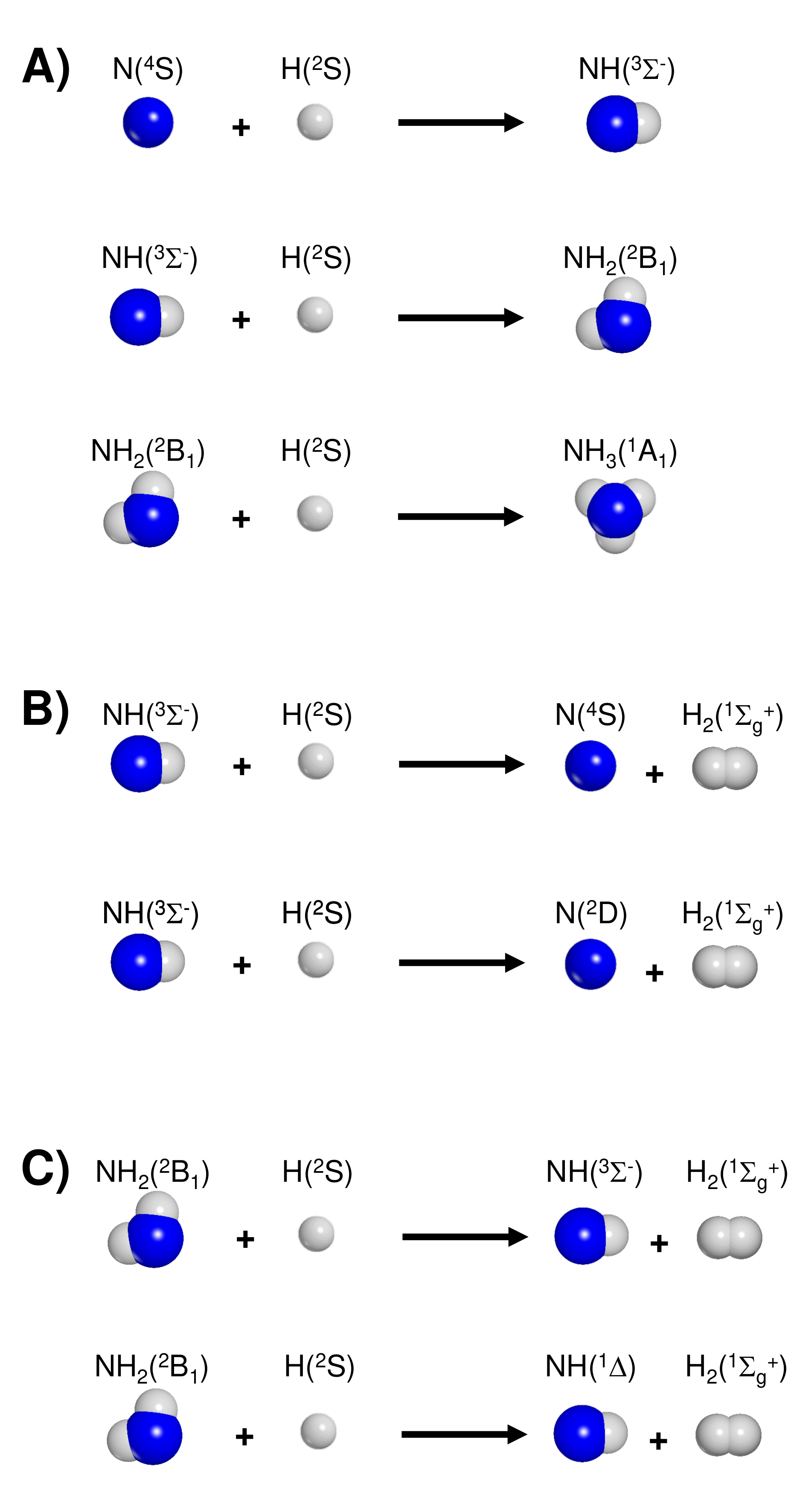}
    \caption{Scheme of the sequential H-addition (A) and of the H-abstraction (B and C) reactions.}
    \label{fig:Scheme of the sequential H-addition (A) and of the H-abstraction (B and C) reactions. The electronic states are also shown in parenthesis.}
    \label{fig:2}
\end{figure}

Previous to the AIMD simulations, we first carried out static calculations in order to elucidate the energetics of the processes under study. The reaction steps towards NH$_3$ formation and the competitive channels are shown in Figure \ref{fig:2}.

Hydrogenation of N to form NH$_3$ involves three steps (see Figure \ref{fig:2}A), leading to the formation of NH, NH$_2$ and NH$_3$. Figure \ref{fig:2}A also shows the electronic ground states of the involved species in the gas phase, i.e., N($^{4}$S), NH($^{3}\Sigma^{-1}$), NH$_2$($^{2}$B$_{1}$) and NH$_3$($^{1}$A$_{1}$). On the water ice surface models, these electronic ground states are kept as such (i.e., no electronic inter-system crossings take place upon adsorption). Hydrogenation reactions, however, have competitive processes, i.e., H-abstraction, in which an incoming H atom abstracts an H atom belonging to the nitrogenated species, reverting the hydrogenation reaction and forming H$_{2}$. The possible H-abstraction reactions are shown in Figure \ref{fig:2}B and \ref{fig:2}C. Figure \ref{fig:2}B represents the H-abstraction on NH to form N and H$_2$. The resulting N can be either in the $^{4}$S ground electronic state or in the $^{2}$D first excited electronic state. Figure \ref{fig:2}C represents the H-abstraction of NH$_2$ to form NH and H$_2$, in which the formed NH can be either in the $^{3}\Sigma^{-1}$ ground electronic state or in the $^{1}\Delta$ first excited electronic state (Figure \ref{fig:2}C). All these possible situations have been considered in this work. 

Since the amount of energy to dissipate is, for the scope of this work, a fundamental quantity, we computed the reaction energies of the three hydrogenation steps in the gas phase at the PBE-D3(BJ) level, using CP2K and Orca. Results were compared with the reference CCSD(T)-F12 values, and, to get deeper insights into the comparison, also with those obtained at $\omega$B97x-D3(BJ). Additionally, energy barriers (if existent) of these reactions, as well as the energetics of the H-abstraction processes (see Figure \ref{fig:2}) were also computed.

The three H-addition reactions to form NH$_3$ are barrierless in the gas phase, namely, the reactants spontaneously evolve towards the products during the optimization, because they involve radical-radical coupling processes and, thus, are driven by spin couplings. Computed reaction energies (reported in Table \ref{tab:1}) are all negative and very large (ranging from ca. -360 and -460 kJ mol$^{-1}$), in which, moreover, the more hydrogenated the product, the more favorable the reaction energy. There is a good agreement among the values computed with all the quantum chemical methods employed and the two codes, the unsigned standard errors being 1--11\% (using CCSD(T)-F12 as the reference method). Results indicate that PBE-D3(BJ) level of theory is accurate enough to describe the energetics of the H-additions. 

\begin{table*}[htp]
    \caption{Computed reaction energies ($\Delta E_{rx}$) for the sequential H-addition reactions in the gas phase using different quantum chemical methods (PBE-D3(BJ), $\omega$B97x-D3(BJ) and single points at CCSD(T)-F12) and codes (CP2K and Orca). All the H-addition reactions have been found to be barrierless, namely, the products are spontaneously formed during the optimization process. Units are in kJ mol$^{-1}$.}

    \centering
    \begin{tabular}{cccccc}
    \toprule
    & CP2K & \multicolumn{3}{c}{Orca}     \\
      \cmidrule(l){3-6}
       $\Delta$E$_{rx}$    &  PBE-D3(BJ)   &  PBE-D3(BJ) & CCSD(T)-F12//PBE-D3(BJ)  & $\omega$B97x-D3(BJ)& CCSD(T)-F12//$\omega$B97x-D3(BJ)\\
       \hline
N + H $\rightarrow$ NH    & -359.3  & -367.2 & -330.8 & -351.5 & -331.2 \\
NH + H $\rightarrow$ NH$_2$ & -406.5  & -415.4 & -399.7 & -411.5 & -399.9 \\
NH$_2$ + H $\rightarrow$ NH$_3$ & -463. 6 & -472.3 & -466.2 & -473.4 & -466.0 \\
\toprule
    \end{tabular}
    \label{tab:1}
\end{table*}

In relation to the competitive H-abstraction reactions (results available as Supplementary Material \citep{sferrero_dissipation_NH3}), it is found that those processes leading to products in their electronic ground-state (i.e., N(S) and NH($^{3}\Sigma^{-1}$)) are exoergic because of the favorable energy of formation of the H-H bond. Nevertheless, this is not the case for processes forming excited state products (i.e., N($^{2}$D) and NH($^{1}\Delta$)), which counterbalance the formation energy of the H-H bond resulting in endoergic reactions, not possible at the cold ISM temperatures. Exoergic reactions present energy barriers of about 10-11 kJ mol$^{-1}$ and 26-29 kJ mol$^{-1}$ for the formation of N($^{4}$S) and NH($^{3}\Sigma^{-1}$), respectively (computed at $\omega$B97x-D3(BJ)//$\omega$B97x-D3(BJ) and CCSD(T)-F12//$\omega$B97x-D3(BJ) levels of theory). Thus, we can conclude that H-abstraction reactions cannot be considered as efficient competitive channels to the H-addition ones, since they are either endoergic processes (while H-additions are all largely exoergic) or present high energy barriers for interstellar conditions (while H-additions are all barrierless). Therefore, to deal with the reaction energy dissipation, we will solely consider the H-additions as the main-occurring reaction channels.

\begin{figure}
    \centering
    \includegraphics[width=\linewidth]{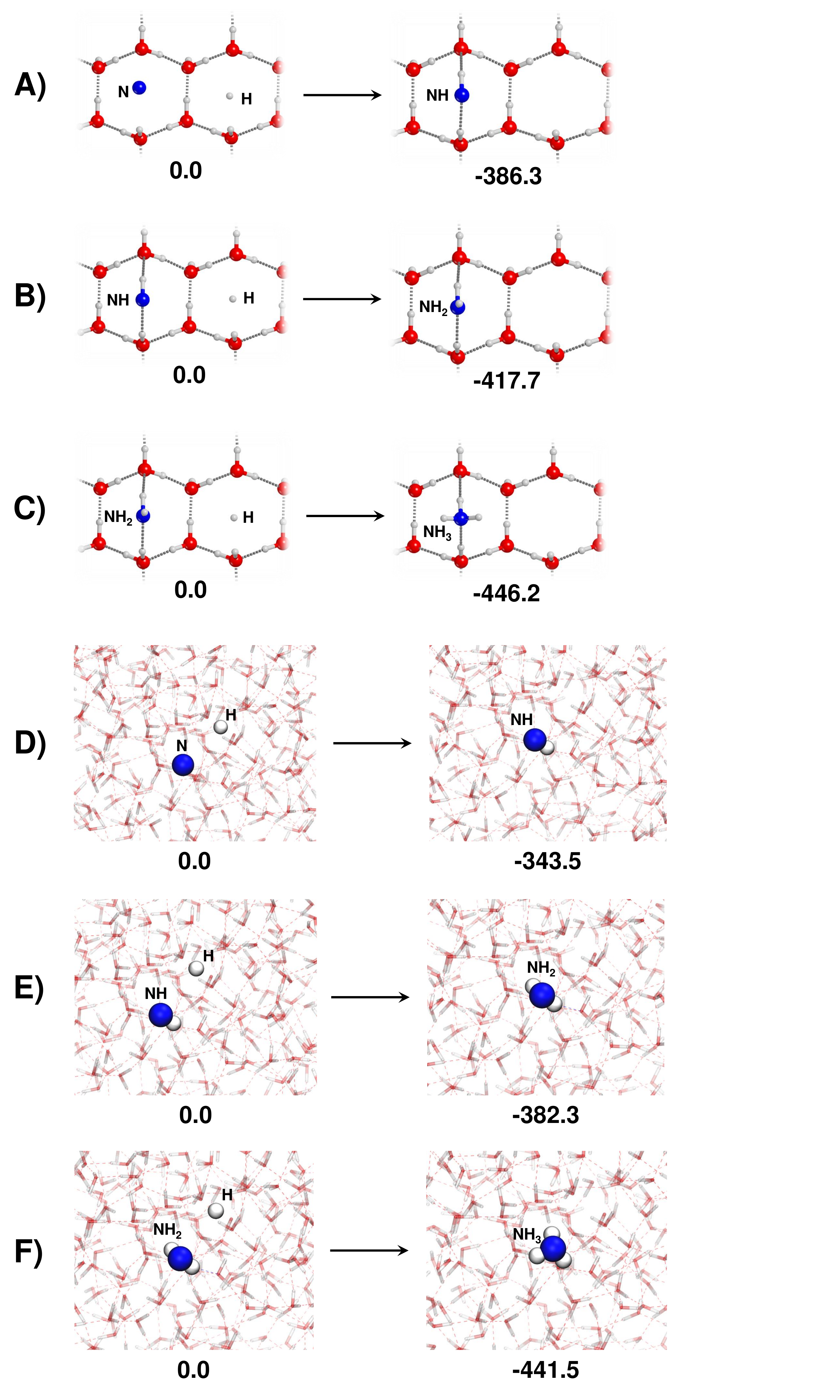}
    \caption{H-addition reactions on the crystalline (A – C) and amorphous (D – F) surfaces. PBE-D3(BJ)-computed reaction energies are also shown, in kJ mol$^{-1}$}
    \label{fig:3}
\end{figure}

On the water ice surfaces, we studied the reactions according to a Langmuir-Hinshelwood (LH) mechanism, namely the reactive species are on the surface and diffuse until they encounter each other. To study the energetics of the reactions, we make here an important assumption: the diffusion of the species on the ice surface already happened. Accordingly, the initial states involve the two reactants in proximity ready to react. Thus, we performed static optimizations on the ice surface models, placing the reactants in two adjacent adsorption sites (i.e., distances of 4.0-4.3 Å on the crystalline surface and of 3.6-4.0 Å on the amorphous one), as depicted in Figure \ref{fig:3}. We optimized first these pre-reactant species in the electronic states that prevent the occurrence of the H-addition reactions due to the Pauli/exclusion repulsion principle (namely, quintet, quadruplet, and triplet for the first, second and third H-addition, respectively), this way obtaining an optimized geometry of these complexes based on only species/surface interactions. From these pre-reactant adducts, then optimizations at the reactive electronic states (namely, triplet, doublet and singlet, respectively) were performed. It is worth mentioning that, at variance to the gas phase, on the surface, the reactant/surface interactions must be partly broken as the reaction takes place. Despite this, our results indicate that, also on the icy surfaces, all the H-addition reactions occur spontaneously during the optimization processes and, accordingly, they are also barrierless processes. To ensure that this chemical behavior is kept also when dynamic effects are accounted for, for each reaction, we run a short AIMD simulation at 10 K, and observed that reactants also spontaneously evolve to products in the first femtoseconds.

Figure \ref{fig:3} also shows the calculated reaction energies of the processes. On the surfaces, similarly to the gas-phase, the reactions are exoergic (between -340 and -440 kJ mol$^{-1}$, values obtained taking as reference the pre-reactant complexes in their non-reactive electronic states). Those reactions occurring on the amorphous model are slightly less negative than on the crystalline model (see Figure \ref{fig:3}) due to the different number of hydrogen bonds established by the adsorbed species with the different surfaces. The trend in which at each H-addition step the reaction energy is more favorable is also kept. In view of these results, we therefore can conclude that placing the reactants at nearly 4 Å leads to very favorable H-addition reactions (i.e., barrierless and with large and negative reaction energies) so that the hydrogenation reactions are not energetically hampered. However, it is worth bearing in mind that, depending on the distance between the two reactants as well as the surface morphology, the reactions are limited by diffusion. 

\subsection{Energy dissipation in the H-addition reactions} \label{sec:AIMD}
Here, we describe the results concerning the energy dissipation in the three H-additions to the atomic N that result from the AIMD simulations.

The initial guess structures were the optimized geometries of the pre-reactant complexes in their non-reactive electronic states, in which the reactive species are ca. 3.6--4.0 Å apart (reactant structure of Figure \ref{fig:3}D, E and F). This intermolecular distance is a suitable choice because it is short enough to neglect the diffusion of the species but long enough to include at least 94\% of the reaction energy in the AIMD simulations. The kinetic energy liberated by the reactions is studied by following the evolution of the velocities of the system particles throughout the entire trajectories. This analysis allowed us to monitor the kinetic energy partitioning between the newly formed species and the ice surface, providing unique information on: i) the kinetic energy fraction retained by the product, ii) the amount of energy dissipated through the ice surface phonon modes, iii) the temperature increase of the ice due to this energy dissipation and iv) the possibility of desorption of the product due to retaining a large fraction of the kinetic energy. Table \ref{tab:2} summarizes part of these data for the studied processes.

\begin{table*}
\caption{Data obtained from the (NVE) AIMD simulations for the reactions investigated. First column represents the reactions studied. Second and third columns report the fraction of kinetic energy transferred to the ice and retained by the newly born species (S), respectively. The fourth column shows the variation of the temperature (in K) undergone by the ice at the end of the simulation. The fifth column reports the computed binding energies (in kJ mol$^{-1}$) of the species at the reaction sites. The sixth and seventh columns summarize the translational kinetic energy (TKE) and the TKE along the z-axis (TKE-z) of the newly formed species (in kJ mol$^{-1}$), respectively}    
    \centering
    \begin{tabular}{lllllll}
    \toprule
Reaction &   $\frac{T_{ice}}{T_{ice} +T_{S}}$    &   $\frac{T_{S}}{T_{ice} +T_{S}}$   &     Temperature variation      & BE &  TKE & TKE-z     \\
\hline
N + H $\rightarrow$ NH & 0.76 & 0.24 & 10 $\rightarrow$  21.6 & 17.5       & 2.7 & 1.4 \\
NH + H $\rightarrow$ NH$_2$ & 0.71 & 0.29 & 10 $\rightarrow$  23   & 18.8       & 2.9 & 1.3 \\
NH$_2$ + H $\rightarrow$ NH$_3$ (Pos1)& 0.73 & 0.27 & 10 $\rightarrow$  24.1 & 22.4       & 5.2 & 1.4 \\
NH$_2$ + H $\rightarrow$ NH$_3$ (Pos2)& 0.58 & 0.42 & 10 $\rightarrow$  23.8 & 34.4       & 3.5 & 2.2 \\
NH$_2$ + H $\rightarrow$ NH$_3$ (Pos3)& 0.90 & 0.10 & 10 $\rightarrow$  27.6 & 36.3       & 2.9 & 2.2 \\
\toprule
\end{tabular}
    \label{tab:2}
\end{table*}

 \label{subsec:tables}

\begin{figure}
    \centering
    \includegraphics[width=\linewidth]{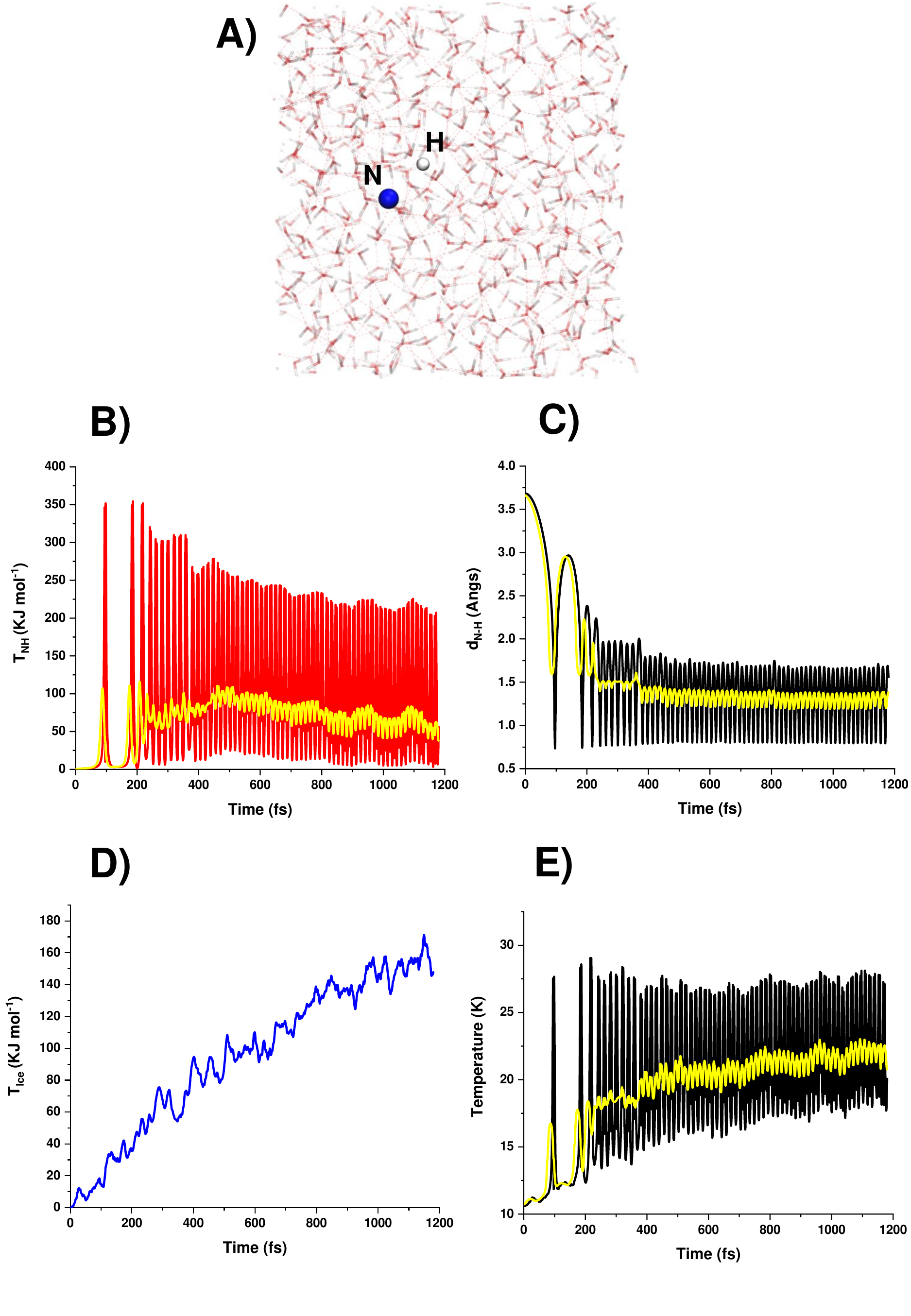}
    \caption{Results of the NVE AIMD simulations for the first H-addition N + H $\rightarrow$ NH: A) initial structure of the simulation; B) evolution of the kinetic energy of NH; C): evolution of the N-H distance; D) evolution of the kinetic energy of the ice; and E) evolution of the temperature of the ice. Instantaneous and averaged values (in yellow) are reported.}
    \label{fig:4}
\end{figure}

According to the static calculations, the first H-addition is the less exothermic step, with a reaction energy of -343.5 kJ mol$^{-1}$ (see Figure \ref{fig:3}D), a considerable amount of energy, whose distribution has been analyzed by following the kinetic energy as depicted in Figure \ref{fig:4}.

Panel A shows the initial positions of N and H on the amorphous surface, panel B the evolution of the kinetic energy of the newly formed NH molecule, panel C the evolution of the N—H distance (starting at ca. 3.6 Å), panel D the evolution of the kinetic energy of the ice surface, and panel E the temperature variation of the ice along the simulation. Figure \ref{fig:4}B shows initial peaks reaching $\approx$350 kJ mol$^{-1}$, which is very close to the NH reaction formation energy on the ice surface, indicating that this is the actual amount of energy released by the reaction. At the beginning of the simulation, the NH species forms in a highly excited vibrational state, as indicated by the high and narrow energy spikes in Figure \ref{fig:4}B. Such transient excited states induce large oscillations of the forming N-H bond (see Figure \ref{fig:4}C). Nevertheless, already in the first femtoseconds, the water ice molecules close to the reaction site partly absorb the reaction energy, a total of about 157 kJ mol$^{-1}$ during the whole simulation (see Figure \ref{fig:4}D). This energy transfer induces a temperature increase of the ice from 10 K to an average of 22 K (see Figure \ref{fig:4}E). Despite this thermal relaxation of the NH molecule by exchanging energy with the surface, it also retains a certain amount of kinetic energy, about 50 kJ mol$^{-1}$ (see average line of Figure \ref{fig:4}B). At the end of the simulation, the kinetic energy liberated by the reaction is distributed as follows: 76\% is transferred to the ice, and 24\%  is retained by the product (see Table \ref{tab:2}). In the Supplementary Material \citep{sferrero_dissipation_NH3}, the evolution along the AIMD simulation of the most relevant energetic components, i.e., the total energy (E$_{TOT}$, potential + kinetic), the potential energy (V$_{TOT}$) and the kinetic energy (T$_{TOT}$) are compared, showing the energy conservation of our AIMD simulations in consistency with the NVE ensemble. 

\begin{figure}
    \centering
    \includegraphics[width=\linewidth]{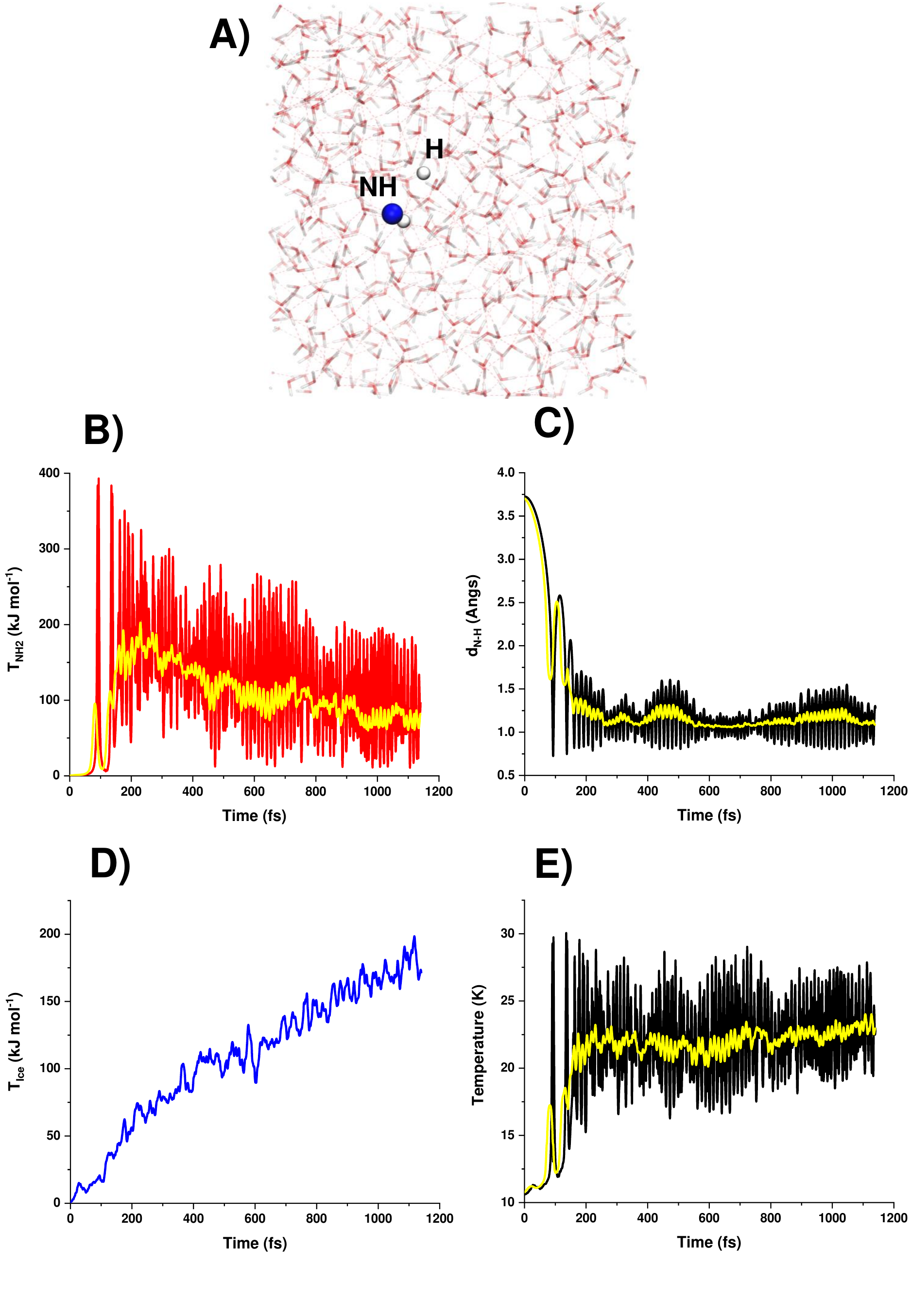}
    \caption{Results of the NVE AIMD simulations for the second H-addition NH + H $\rightarrow$ NH$_2$: A) initial structure of the simulation; B) evolution of the kinetic energy of NH$_2$; C): evolution of the newly formed N-H bond distance; D) evolution of the kinetic energy of the ice; and E) evolution of the temperature of the ice. Instantaneous and averaged values (in yellow) are reported.}
    \label{fig:5}
\end{figure}

Results of the second H-addition, i.e., NH + H $\rightarrow$ NH$_2$, are shown in Figure \ref{fig:5}. The first peak of the kinetic energy of the newly formed NH$_2$ (Figure \ref{fig:5}B) is in good agreement with the reaction energy of the process (-382 kJ mol$^{-1}$). This second hydrogenation shows a faster relaxation in comparison with the first one, as deduced from the narrower oscillations in the kinetic energy of NH$_2$ (Figure \ref{fig:5}B) and from the lower oscillations of the formed N-H bond distance (Figure \ref{fig:5}C). The reason for this faster relaxation is that NH$_2$ has more internal degrees of freedom than NH, which can strongly couple with the newly formed N-H bond as well as with the vibrations of the icy water molecules of the surface. As the released energy for this reaction is larger than the first H-addition, more kinetic energy is dissipated thorough the ice surface, about 180 kJ mol$^{-1}$ (see Figure \ref{fig:5}D), leading to a temperature increase of the ice up to 23 K (see Figure \ref{fig:5}E), and the internal kinetic energy retained by the newly formed NH$_2$ molecule is of about 73 kJ mol$^{-1}$ (see average line of Figure \ref{fig:5}B), resulting in a distribution of 71\% and 29\%, respectively (see Table \ref{tab:2}).

\begin{figure}
    \centering
    \includegraphics[width=\linewidth]{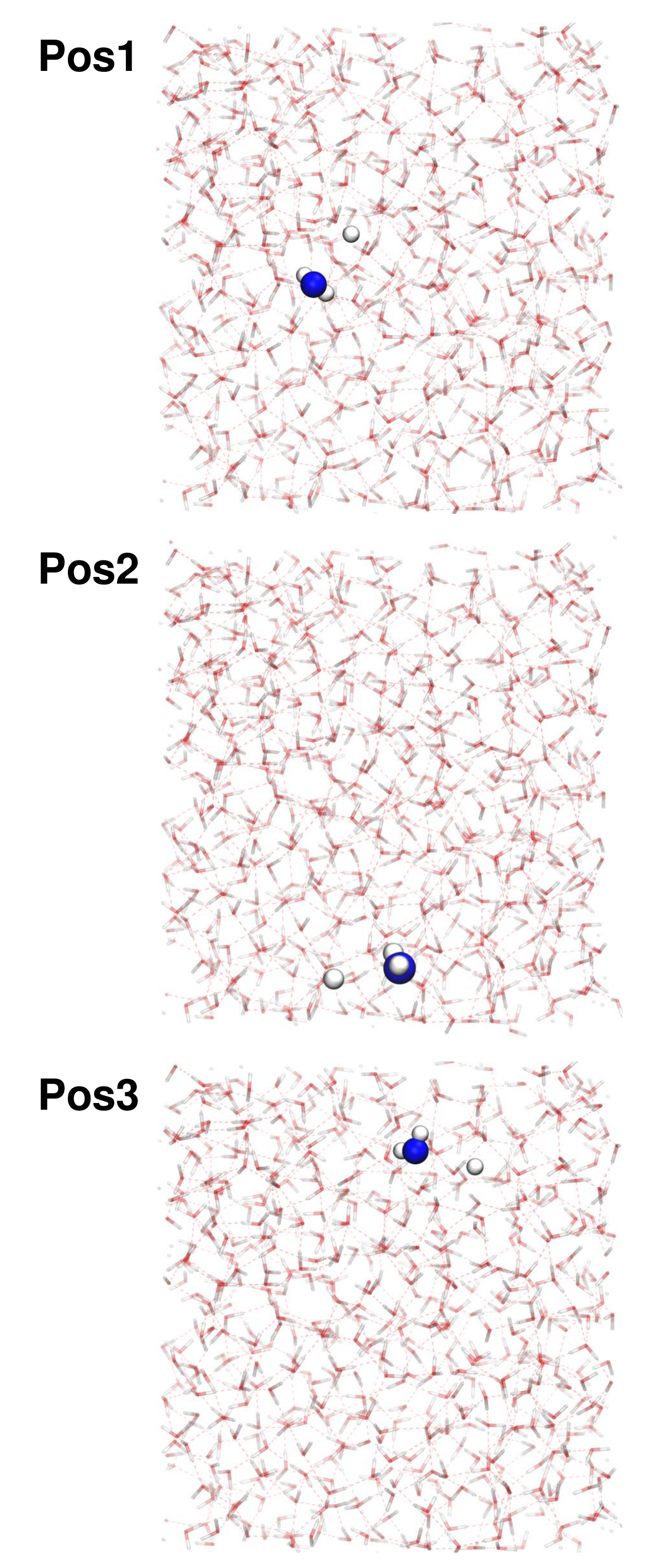}
    \caption{Starting positions of the reactive species for the NVE AIMD simulations relative to the third H-addition NH$_2$ + H $\rightarrow$ NH$_3$}
    \label{fig:6}
\end{figure}

The third H-addition reaction (i.e., NH$_2$ + H $\rightarrow$ NH$_3$) is that releasing the largest amount of energy ($\approx$-440 kJ mol$^{-1}$). Because of that, and also due to the fact that this is the last step to form the final NH$_3$ product, for this reaction we calculated the trajectories from three different starting positions of the reactants, with the aim to consider different possible situations as far as the surface morphology is concerned. The chosen initial positions are shown in Figure \ref{fig:6} and present the following features: i) Pos1 (the same initial positions as the first and second H-additions), in which NH$_2$ and H are placed in an outermost position of the surface (i.e., not inside a cavity) and NH$_2$ acts as an H-bond donor; ii) Pos2, in which NH$_2$ and H are placed in an outermost position of the surface and NH$_2$ acts as an H-bond acceptor; iii) Pos3, in which NH$_2$ and H are placed within a cavity and NH$_2$ acts as an H-bond donor. By proceeding this way, we roughly sample different sites of the amorphous surface that can likely exhibit specific trends in the energy dissipation, as observed for the H$_2$ formation case \citep{pantaleone2021h2}.

For the AIMD simulation from Pos2, the trends in the energy partition are similar to those observed for the other H-additions. Graphs on the evolution of the kinetic energies, oscillation of the newly N-H bond formed, and temperature variation of the ice are shown in the Supplementary Material \citep{sferrero_dissipation_NH3}.
The NH$_3$ molecule remains highly excited during the whole simulation which is reflected by large oscillations of both the kinetic energy and the N—H bond length. The amount of the reaction energy that is transferred to the surface as kinetic energy is about 159 kJ mol$^{-1}$, resulting in a temperature increase of the ice from 10 to 24 K along the simulation. In contrast, in the simulated trajectories starting from Pos1 and Pos3, one can distinguish a different energy dissipation behavior, as provided by the graphs shown in Figure \ref{fig:7}A and E (for Pos1 and Pos3, respectively): the newly formed NH$_3$ molecule exchanges more energy with the surface and in shorter timescales. This can be seen from the abrupt change in the oscillation amplitude of the kinetic energy of NH$_3$, and from the ice temperature variation, while there is a sudden increase in the evolution of the kinetic energy of the ice at these simulation times. For Pos1, the kinetic energy of NH$_3$ is about 78 kJ mol$^{-1}$ while that transferred to the surface is 210 kJ mol$^{-1}$, and for Pos3 they are 34 kJ mol$^{-1}$ and 321 kJ mol$^{-1}$, indicating that the amount of the nascent energy transferred to the surface is larger in Pos1 and Pos3 than in Pos2. Interestingly, in Pos1 and Pos3, the formation of an ion pair transient species of the kind H$_3$O$^{+}$/NH$_{2}^{-}$ at around 500 and 200 fs, respectively, takes place, which is crucial to efficiently dissipate the large amount of reaction energy. The occurrence of this event (i.e., formation of a transient species) and its relevance in the energy dissipation is addressed in the next section.

\section{Discussion} \label{sec:discussion}

\subsection{Energy dissipation, chemical desorption and diffusion}

An important aspect to consider here is the possibility that chemical desorption (CD) of the newly formed products can occur after the reaction. A plausible CD mechanism is that the newly formed species desorbs because it conserves part of the reaction energy in the form of translational energy, which is used to desorb towards the gas phase. To assess this point, we extracted the translational kinetic energy (TKE) of the newly formed species as the kinetic energy associated with the center of mass \citep{takahashi1999product}. The TKE component along the axis normal to the surface (in this case the z-axis) is the component that would bring the newly formed species towards the gas phase. Therefore, the z component of the TKE (TKE-z) was extracted and compared to the BEs of the newly formed species (see Table \ref{tab:2}). We find out that, in all cases, TKE-z contribution accounts for just a small fraction of the BE (between 4.3--8.0\%) and, accordingly, chemical desorption is not expected. However, it could also be possible that the newly formed species, while diffusing on the ice surface once formed, can stumble over a surface irregularity (e.g. dangling H atoms) and, if it was so, the total TKE could be canalized into the desorption direction. This was indeed the case of H$_2$ formation,29 which was ejected into the gas-phase upon stumbling over an outermost water molecule of the surface. Accordingly, to assess this point, we compared the total TKE retained by the born species with their BEs. Like in TKE-z, none of the total TKE values are larger than the computed BEs (total TKEs are between 8--16\% of the corresponding BEs), so we can conclude, again, that chemical desorption is not expected to take place through this mechanism. Therefore, the formed species are doomed to remain attached to the ice surface.

On the other hand, it is worth mentioning that the timescale in which most of the energy dissipates throughout the surface takes place in about 1 ps. However, the H-addition reactions, as they are barrierless, are controlled by the diffusion of the H atoms. It is thus interesting to assess if the H diffusion can compete with the energy dissipation because, in case of similar timescales, the intermediate H-addition reactions (i.e., the second and the third hydrogenations) can take place with the newly formed species in excited vibrational states. To check qualitatively upon this point, we have proceeded as follows. It is estimated in the literature that the H diffusion on water ice can be 5.8x10$^{-1}$ cm$^2$ s$^{-1}$ or 3.3x10$^{-7}$ cm$^2$ s$^{-1}$ at 25 K \citep{asgeirsson2017long}, depending on the presence/absence of inert species blocking the most stable adsorption sites, and considering that these numbers are upper limits as they assume that there is always an H atom ready to diffuse, omitting the probability that an H atom lands on the surface \citep{watanabe2010direct, hama2012mechanism}. According to these diffusion coefficients, an H atom scans a surface area of 745.6 Å$^2$ (the area of our unit cell surfaces) in 1.28 ms and 0.89 ns, respectively, which are significantly longer than our energy dissipation timescale (1 ps). This supports the idea that the formation of NH$_3$ by hydrogenation of N is a true stepwise H-addition process, in which the NH formation energy is efficiently dissipated throughout the grain while the newly formed species get relaxed from any vibrationally excited states before entering the next hydrogenation step.

The conclusions described above, i.e., the non-occurrence of chemical desorption and that the energy relaxation of the newly formed species is faster than the H diffusion, are very important aspects for the interstellar chemistry. That is, the NH, NH$_2$ and NH$_3$ species are not ejected into the gas phase once formed, thereby ruling out a direct CD mechanism. After each hydrogenation step, the associated reaction energies are promptly dissipated, and the newly formed species remain in low vibrational states, ready for a new hydrogenation (unlike if they become desorbed, which, to be the case, would be in excited vibrational states, presenting a different reactivity compared to their fundamental states). Remarkably, this picture is consistent with the fact that NH$_3$ is a component of the interstellar ices: after the completion of the full H-addition sequence, NH$_3$ remains on the water ice surface, which upon subsequent H$_2$O accretion and/or in situ formation on the grains can become a bulky component of the ice mantles, in agreement with the observational IR features of interstellar ices \citep{boogert2015observations}. However, it is worth mentioning that our simulations do not account for additional surface phenomena that can induce CD. This would be, for instance, the occurrence of other large exothermic reactions (e.g., the H$_2$ formation) nearby the newly formed species, which can produce a local heating in their surroundings triggering the desorption into the gas. In fact, this type of indirect CD is usually included in the astrochemical models and allows explaining the gas-phase abundances of relevant interstellar molecules (e.g., H$_2$O, CO, H$_2$CO, CH$_3$OH and several iCOMs) \citep{vasyunin2013reactive, garrod2007non,ruaud2015modelling,hocuk2015interplay, minissale2016dust}, by considering that a few per cent of these components, formed on the grain surfaces, are ejected into the gas-phase through this non-thermal mechanism. Interestingly, direct chemical desorption is also invoked to occur through H-abstraction reactions. This has experimentally been investigated for the CO and H$_2$CO cases \citep{minissale2016hydrogenation}, in which authors found that for the [CO + H] and [H$_2$CO + H] reactive systems, chemical desorption of CO and H$_2$CO was detected, presumably due to the ease of reaction and large exothermicity of the HCO + H $\rightarrow$ CO + H$_2$ and CH$_3$O/H$_2$COH + H $\rightarrow$ H$_2$CO + H$_2$ H-abstraction processes. Nonetheless, similar chemical desorption is not expected for the N-bearing species dealt in this work because the H-abstraction reactions, as shown above, present a limited exothermicity, rendering unlikely the chemical desorption of these species. 

\begin{figure}
    \centering
    \includegraphics[width=\linewidth]{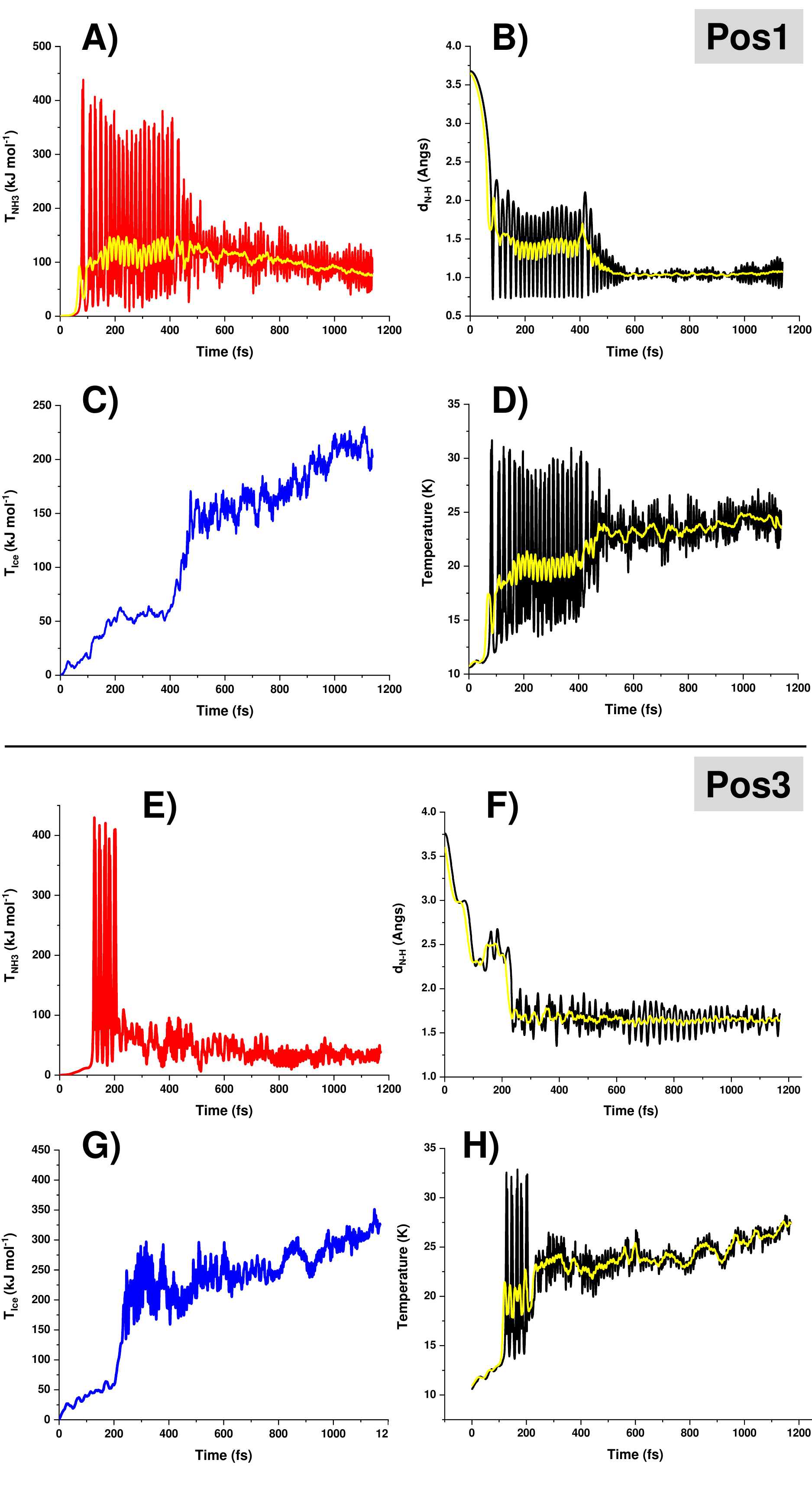}
    \caption{Results of the NVE AIMD simulations for the third H-addition NH$_2$ + H $\rightarrow$ NH$_3$ starting from positions Pos1 and Pos3: evolution of the kinetic energy of NH$_3$ (A and E); evolution of the newly formed N-H bond distance (B and F); evolution of the kinetic energy of the ice (C and G); and evolution of the temperature of the ice (D and H). Instantaneous and averaged values (in yellow) are reported.}
    \label{fig:7}
\end{figure}

\subsection{Energy dissipation mechanisms}
Due to the large amount of energy released by every H-addition step, each product is formed in a highly excited vibrational state, which relaxes to lower vibrational states at different timescales. This is indicative that different dissipation energy mechanisms are involved in the relaxation of the newly born species. 

\begin{figure}
    \centering
    \includegraphics[width=\linewidth]{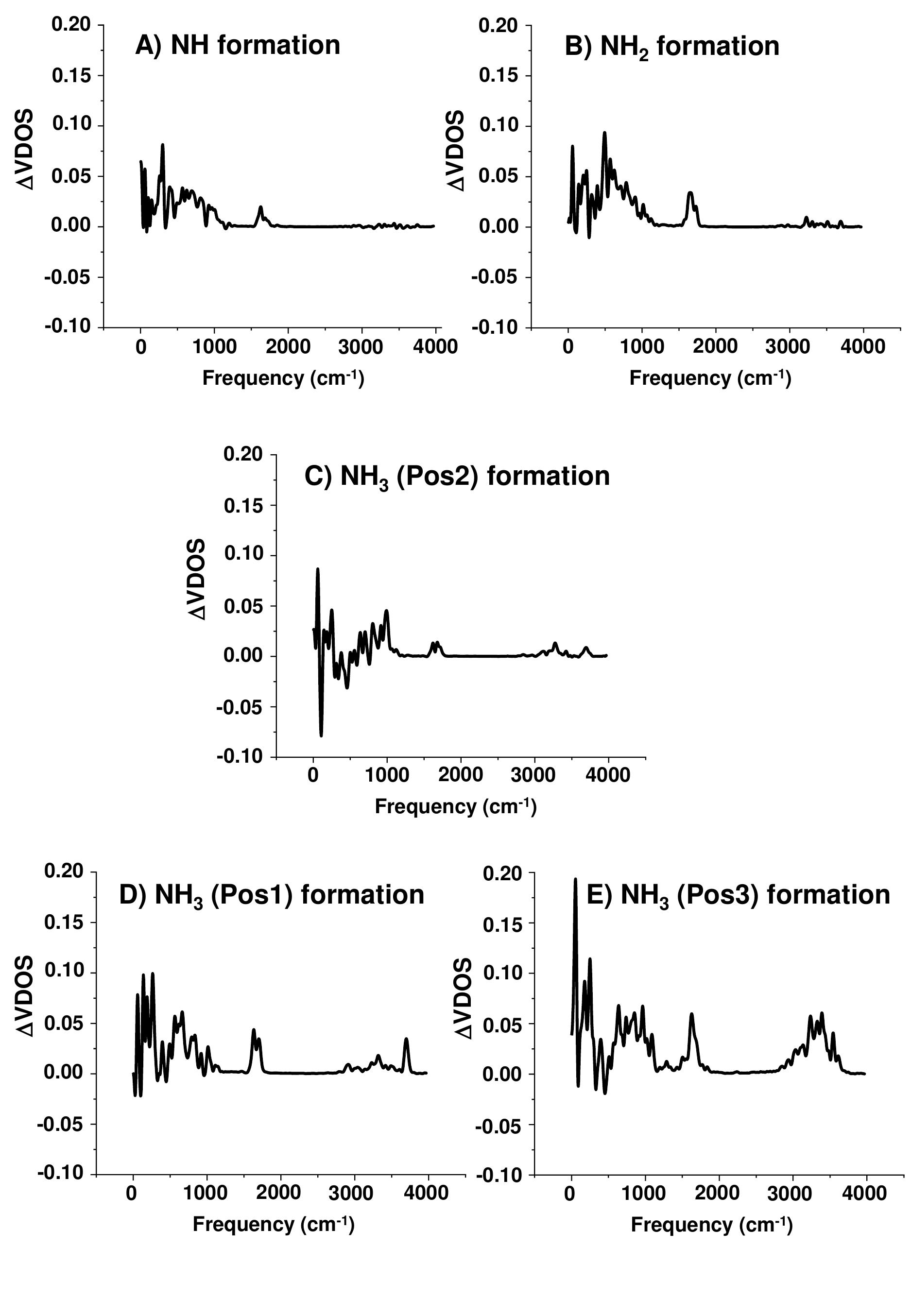}
    \caption{Difference of the simulated VDOS power spectrum ($\Delta$VDOS) between the computed trajectories for the surfaces when the reaction takes place and for the isolated surface.}
    \label{fig:8}
\end{figure}

Upon formation, the vibrational modes of the formed species (stretching and bending for NH$_2$ and NH$_3$, and only stretching for NH) are highly excited, while this is not the case for the 10 K-thermalized internal modes of the icy water molecules (consisting of stretching and bending vibrations of individual water molecules and the low-energy libration modes arising from the lattice nature of the water ice surface). Accordingly, at first instance, there is a coupling between the excited stretching and bending modes of the newly formed species and the libration modes of the underneath water ice surface. To have evidence on that, we calculated the trajectory VDOS power spectrum (which highlights the active vibrational modes during the simulations) for the water ice surface when the reactions occur and for the isolated water ice surface thermalized at 10 K, and we plot the difference of the computed VDOS power spectra ($\Delta$VDOS) to highlight the difference in the population of the vibrational modes. Figure \ref{fig:8}A, B and C shows the $\Delta$VDOS plot during the reactions involving the formation of NH, NH$_2$ and NH$_3$ (Pos2), while in Figure \ref{fig:8}D and E the same but for NH$_3$ formation from Pos1 and Pos3. In all the cases, the bands that become more populated correspond to the libration frequencies. Interestingly, for the formation of NH, NH$_2$ and NH$_3$ (Pos2), the water ice bending modes become populated, whereas the stretching remain unpopulated, indicating a vibrational coupling between the bending modes of the newly formed species and the ones of the water molecules in the ice surface. These results, thus, clearly indicate that a main mechanism responsible for the dissipation of the large excess of the reaction energy is the intermolecular vibrational coupling between the vibrational modes of the born species and the libration modes (with minor contribution to the bending modes) of the water ice. 

\begin{figure}
    \centering
    \includegraphics[width=\linewidth]{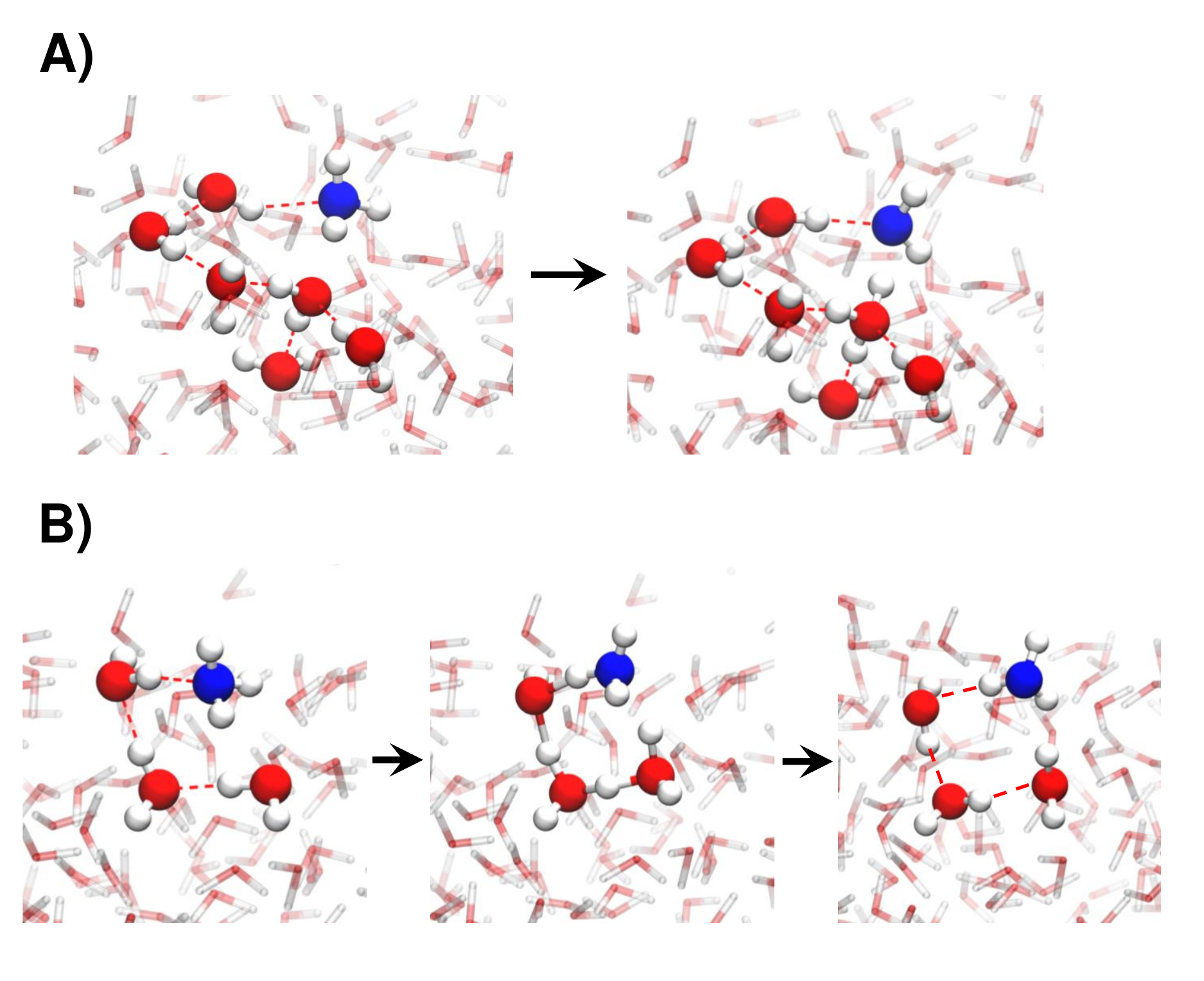}
    \caption{Snapshots of the trajectories relative to the NH$_3$ formation (A: Pos1; B: Pos3), in which the transient H$_3$O$^{+}$/NH$_{2}^{-}$ ion pair species forms. In panel A, the ion pair is directly formed. In panel B) a proton transfer relay mechanism takes place.}
    \label{fig:9}
\end{figure}

For reactions represented in Figure \ref{fig:8}D and E, in addition to the libration modes, there is a significant population of the icy water stretching modes and a larger population of the bending modes compared to the $\Delta$VDOS plots of Figure \ref{fig:8}A, B and C. Thus, for these cases an additional energy mechanism operates, which in this case is based on the formation of the transient H$_3$O$^{+}$/NH$_{2}^{-}$ ion pair species (see Figure \ref{fig:9}) at 400 fs and 200 fs during the NH$_3$ formation (Pos1 and Pos3 trajectories, respectively). The ion pair nature of these species is confirmed by their calculated charges: the N charge is lowering from -0.34 to -0.74 and the charge of the oxygen atom involved in the ion pair is changing from -0.50 to -0.38, approximately. These species are formed by a proton transfer from NH$_3$ to a nearby surface H$_2$O molecule. This proton transfer could seem surprising as in water solution the expected ion pair is NH4$^{+}$/OH$^{-}$. Here the opposite reaction to form H$_3$O$^{+}$/NH$_{2}^{-}$ occurs due to the large internal energy of the newly formed NH$_3$ in the early stages of its formation in which the energy of the highly excited N-H stretching modes is enough to transfer the proton towards a nearby water molecule. As the ion pair is formed, a sudden energy transfer occurs towards the water ice surface (see Figure \ref{fig:7}). Formation of these species permit to populate the high-frequency bending and stretching modes of nearby water molecules. In this case, the population is not exclusively due to a vibrational coupling between vibrational modes of the two partners but through a “by-side” chemical reaction (a proton transfer), which injects kinetic energy from the newly formed NH$_3$ to the stretching modes of the icy water molecules. This  “chemical” energy transfer is highlighted in the $\Delta$VDOS plots shown in Figure \ref{fig:8}D and E, where the stretching peak around 3600-3700 cm$^{-1}$ is populated, while completely absent in the spectra of Figure \ref{fig:8}A, B and C, in which the ion pair is not formed. Obviously, the population of these peaks is rather low as the ion pair has a short lifetime (50-100 fs) while the VDOS spectra is for the entire trajectory (1200 fs). Interestingly, the fate of the transient H$_3$O$^{+}$/NH$_{2}^{-}$ ion pair in Pos1 and Pos3 is different, which affects the efficiency of the relaxation mechanism. While in Pos1 the transferred proton reverts directly to NH$_3$ (see Figure \ref{fig:9}A), in Pos3 the proton does not rebind to NH$_{2}^{-}$ but remains on H$_3$O$^{+}$, which starts a proton transfer to the NH$_{2}^{-}$ species (recovering NH$_3$) via a proton shuttle mechanism involving 3 water molecules (see Figure \ref{fig:9}B).

The proton relay mechanism ensures an even more efficient relaxation of the newly formed NH$_3$: the transferred proton, engaging more than one water molecule, allows for a higher degree of energy dissipation towards the icy surface. It is worth mentioning that, due to the high computational cost of these simulations, collecting enough trajectories to arrive at a robust statistical picture on the occurrence of this “chemical” energy dissipation mechanisms would be overwhelming. Therefore, with the simulated trajectories, we can have hints on the process: i) the formation of the transient ion pairs is geometrically constrained; it is only possible if the proton belonging to the N-H vibrationally excited bond is close enough to the lone pair of an O atom of the ice surface acting as a proton acceptor, and ii) the formation of the ion pair is more probable from NH$_3$ than from NH$_2$ and NH in agreement with their deprotonation energy (NH$_3$ $<$ NH$_2$ $<$ NH). Despite this, we cannot exclude the occurrence of this mechanism also during the NH and NH$_2$ formation, since it was also observed in the H$_2$ formation \citep{pantaleone2021h2}. 

Finally, we would like to stress that we are not including nuclear quantum effects (NQEs) \citep{markland2018nuclear, ceriotti2016nuclear, huppert2022simulation}, which can be of great significance for hydrogen bonded systems at these very low temperatures. NQEs can, indeed, play a crucial role in both the energy transmission and the energy dissipation processes. However, considering the size of the systems investigated, inclusion of NQEs in our simulations is computationally almost out of reach and unpractical. Along this line, efforts dedicated to alleviate the zero-point energy (ZPE) leakage problem in classical trajectory calculations are ongoing \citep{habershon2009zero, mukherjee2022hessian, buchholz2018herman, mangaud2019fluctuation, brieuc2016zero}.

\subsection{Astrophysical implications}
\begin{figure}
    \centering
    \includegraphics[width=\linewidth]{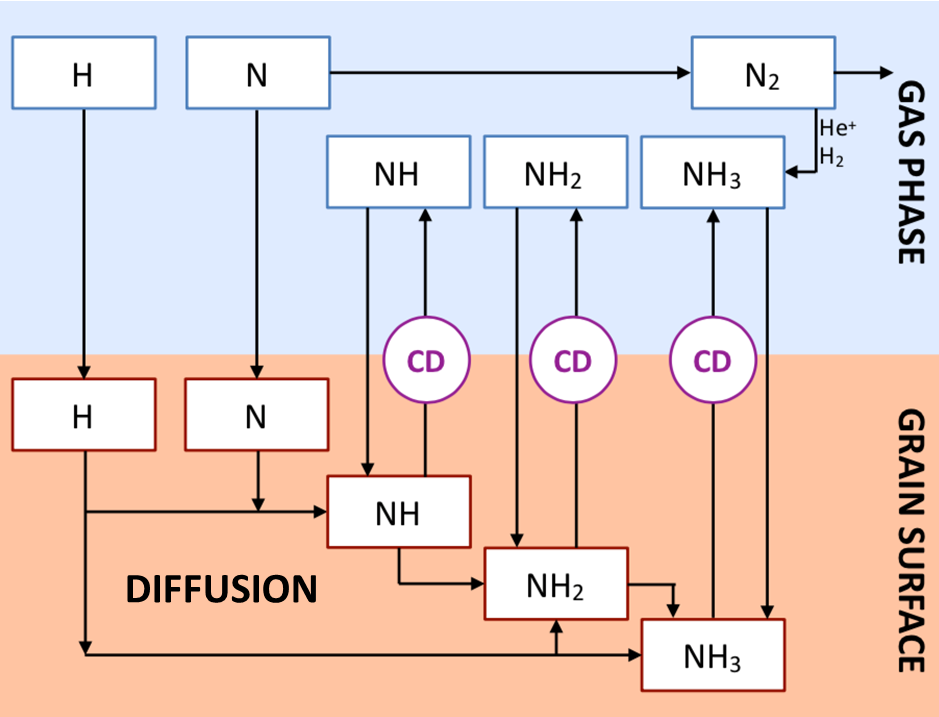}
    \caption{Scheme of the ammonia formation in cold molecular clouds (adapted from \citet{de2022tracking, tinacci2022theoretical}). The major process forming ammonia is the hydrogenation of frozen N atoms into NH, followed by NH$_2$ and finally NH$_3$. Ammonia formed in this way, on the grain surfaces, remains frozen onto the grain mantles, because its binding energy is too large for the thermal sublimation to be efficient. Only the fraction released by the chemical desorption (CD) would be gaseous ammonia.}
    \label{fig:10}
\end{figure}

Ammonia is a very important molecule in Astrochemistry, as it is a major reservoir of nitrogen in the molecular regions of the ISM. Its chemistry has been studied by various authors and a simple scheme has been shown in \citet{tinacci2022theoretical,de2022tracking}, from which the present Figure \ref{fig:10}, describing the ammonia formation in cold molecular clouds, has been adapted for helping the discussion here. Briefly, ammonia is prevalently formed at the beginning of the evolution from a translucent to a molecular cloud (e.g. \citet{ceccarelli2022organic}) by the addition of H-atoms to the N-atoms frozen onto the interstellar grain surfaces, the process studied in this work (Figure \ref{fig:10}). The hydrogenation of frozen nitrogen atoms continues also during the molecular cloud phase, but with smaller efficiency as the remaining gaseous nitrogen becomes incorporated into molecular nitrogen and some of it goes into ammonia via gas-phase reactions. In cold molecular clouds, the only mechanism to eject the frozen ammonia into the gas-phase is via the chemical desorption during the N hydrogenation process because the temperature of cold molecular clouds ($\sim$10 K) is too low with respect to the ammonia binding energy for thermal sublimation being efficient \citep{tinacci2022theoretical}.

In practice, the bulk of ammonia observed in the ISM, when considering the frozen and gaseous form altogether, is built on the grain surfaces at early times and it is mainly due to the hydrogenation of nitrogen atoms on the grain surfaces. Please note that frozen water coats very quickly the bare silicate/carbonaceous surface \citep{tinacci2022theoretical}, so that the nitrogen hydrogenation mostly occurs on icy surfaces. If the hydrogenation process were counterbalanced by the H-extraction competition or the dissipation of the chemical energy were inefficient, making possible a large degree of chemical desorption, the amount of formed ammonia would be reduced. Our computations show, on the contrary, that the H-abstraction processes are completely negligible, and that direct chemical desorption is an extremely improbable process, supporting the assumptions of the present astrochemical models that are based on indirect chemical desorption.

\section{Conclusions}
In this paper, the formation of interstellar NH$_3$ through the successive hydrogenation of atomic N, i.e., N($^4$S) + H($^2$S) $\rightarrow$ NH($^3\Sigma^-$) + H($^2$S) $\rightarrow$ NH$_2$($^2$B$_1$) + H($^2$S) $\rightarrow$ NH$_3$($^1$A$_1$), on water ice surfaces has been investigated at a quantum mechanical DFT PBE-D3(BJ) level, paying special attention to the third body effect exerted by the ice as energy sink and the mechanisms through which the reaction energies dissipate throughout the water ice surface. 
Static calculations indicate that the H-addition reactions leading to the final NH$_3$ are barrierless, both on crystalline and amorphous surfaces, and exhibit large and very negative reaction energies (between -340 and -440 kJ mol-1). The alternative H-abstraction reactions, i.e., NH + H $\rightarrow$ N + H$_2$ and NH$_2$ + H $\rightarrow$ NH + H$_2$, do not compete with the NH$_3$ formation.
Ab initio molecular dynamics (AIMD) simulations within the microcanonical NVE ensemble have been performed to investigate the fate of the energy released in each hydrogenation step and the mechanisms through which this energy flows among the different parts of the system. This has been done by monitoring the evolution over time of the most relevant energetic components: the potential energy, and the kinetic energies of the newly formed species and the ice. AIMD simulations were centred only for the reactions occurring on the amorphous water ice surface model. Simulation results show that about 58\%-90\% of the energy liberated by the reaction is quickly absorbed (within 1 ps) by the ice, which, as a result, is heated up by about 12 – 18 K. Since the amount of translational kinetic energy retained by the newly formed species is small compared to their binding energies, the products are doomed to remain adsorbed on the surface without being released to the gas phase, that is, chemical desorption is not expected. Moreover, we estimated that the velocity of the energy dissipation is much faster than the H surface diffusion, indicating that the NH$_3$ formation through the H-addition adopts a true three-step elementary mechanism, in which, for a given H-addition, the energy is dissipated before the next H-addition. According to these results, we can figure out that the formation of NH$_3$ takes place through a continuous H-addition until reaching the formation of the final product, NH$_3$, which remains on the water ice, becoming over time a bulky component of the interstellar ice, in agreement with observational evidence.

Finally, by computing the vibrational density of states of the water molecules of the ice surface along the AIMD trajectories, we elucidated the mechanisms through which the energy transfer from the formed species to the ice surface takes place. We identified a general channel based on the vibrational coupling between the highly excited stretching and bending vibrational modes of the newly formed species and the low-energy libration modes of the water ice surface. Additionally, a second and more exclusive mechanism (observed in the NH$_3$ formation trajectories but not in the NH and NH$_2$ ones) has been found. It is based on the formation of a transient H$_3$O$^{+}$/NH$_{2}^{-}$ ion pair due to a proton transfer from NH$_3$ to a nearby icy H$_2$O molecule, which permits a faster energy dissipation, thereby relaxing more efficiently the newly born NH$_3$.

\section{acknowledgments}
This project has received funding within the European Union’s Horizon 2020 research and innovation programme from the European Research Council (ERC) for the projects ``The Dawn of Organic Chemistry” (DOC), grant agreement No 741002 and ``Quantum Chemistry on Interstellar Grains” (QUANTUMGRAIN), grant agreement No 865657, and from the Marie Sklodowska-Curie for the project ``Astro-Chemical Origins” (ACO), grant agreement No 811312. S.F. wishes to thank Lorenzo Tinacci for the useful and inspiring discussions on the subject. Supplementary Material consisting of (i) the energetics of the gas-phase reactions for both the H-additions and H-abstractions, (ii) the evolution with time of the total, potential and kinetic energies of the studied processes, and (iii) results of the NVE AIMD simulations for the NH$_3$ formation from Pos2, is available at \href{https://doi.org/10.5281/zenodo.7115984}{ESI} \citep{sferrero_dissipation_NH3}.


\bibliography{bib}{}

\begin{thebibliography}{}
\expandafter\ifx\csname natexlab\endcsname\relax\def\natexlab#1{#1}\fi
\providecommand{\url}[1]{\href{#1}{#1}}
\providecommand{\dodoi}[1]{doi:~\href{http://doi.org/#1}{\nolinkurl{#1}}}
\providecommand{\doeprint}[1]{\href{http://ascl.net/#1}{\nolinkurl{http://ascl.net/#1}}}
\providecommand{\doarXiv}[1]{\href{https://arxiv.org/abs/#1}{\nolinkurl{https://arxiv.org/abs/#1}}}

\bibitem[{{\'A}sgeirsson {et~al.}(2017){\'A}sgeirsson, J{\'o}nsson, \&
  Wikfeldt}]{asgeirsson2017long}
{\'A}sgeirsson, V., J{\'o}nsson, H., \& Wikfeldt, K. 2017, The Journal of
  Physical Chemistry C, 121, 1648,
  \dodoi{https://doi.org/10.1021/acs.jpcc.6b10636}

\bibitem[{Boogert {et~al.}(2015)Boogert, Gerakines, \&
  Whittet}]{boogert2015observations}
Boogert, A.~A., Gerakines, P.~A., \& Whittet, D.~C. 2015, ARA\&A, 53, 541,
  \dodoi{10.1146/annurev-astro-082214-122348}

\bibitem[{Brehm \& Kirchner(2011)}]{brehm2011travis}
Brehm, M., \& Kirchner, B. 2011, J. Chem. Inf. Model., 51, 2007–2023,
  \dodoi{https://doi.org/10.1021/ci200217w}

\bibitem[{Brehm {et~al.}(2020)Brehm, Thomas, Gehrke, \&
  Kirchner}]{brehm2020travis}
Brehm, M., Thomas, M., Gehrke, S., \& Kirchner, B. 2020, J. Chem. Phys., 152,
  164105, \dodoi{https://doi.org/10.1063/5.0005078}

\bibitem[{Brieuc {et~al.}(2016)Brieuc, Bronstein, Dammak, Depondt, Finocchi, \&
  Hayoun}]{brieuc2016zero}
Brieuc, F., Bronstein, Y., Dammak, H., {et~al.} 2016, JCTC, 12, 5688,
  \dodoi{10.1021/acs.jctc.6b00684}

\bibitem[{Buchholz {et~al.}(2018)Buchholz, Fallacara, Gottwald, Ceotto,
  Grossmann, \& Ivanov}]{buchholz2018herman}
Buchholz, M., Fallacara, E., Gottwald, F., {et~al.} 2018, Chem. Phys., 515,
  231, \dodoi{https://doi.org/10.1016/j.chemphys.2018.06.008}

\bibitem[{Bussi {et~al.}(2007)Bussi, Donadio, \&
  Parrinello}]{bussi2007canonical}
Bussi, G., Donadio, D., \& Parrinello, M. 2007, J. Chem. Phys., 126, 014101,
  \dodoi{https://doi.org/10.1063/1.2408420}

\bibitem[{Casassa {et~al.}(1997)Casassa, Ugliengo, \&
  Pisani}]{casassa1997proton}
Casassa, S., Ugliengo, P., \& Pisani, C. 1997, J. Chem. Phys., 106, 8030,
  \dodoi{https://doi.org/10.1063/1.473813}

\bibitem[{Ceccarelli {et~al.}(2017)Ceccarelli, Caselli, Fontani, Neri,
  L{\'o}pez-Sepulcre, Codella, Feng, Jim{\'e}nez-Serra, Lefloch, Pineda,
  {et~al.}}]{ceccarelli2017seeds}
Ceccarelli, C., Caselli, P., Fontani, F., {et~al.} 2017, ApJ, 850, 176,
  \dodoi{10.3847/1538-4357/aa961d}

\bibitem[{Ceccarelli {et~al.}(2022)Ceccarelli, Codella, Balucani,
  Bockel{\'e}e-Morvan, Herbst, Vastel, Caselli, Favre, Lefloch, \&
  {\"O}berg}]{ceccarelli2022organic}
Ceccarelli, C., Codella, C., Balucani, N., {et~al.} 2022, arXiv preprint
  arXiv:2206.13270, \dodoi{https://doi.org/10.48550/arXiv.2206.13270}

\bibitem[{Ceriotti {et~al.}(2016)Ceriotti, Fang, Kusalik, McKenzie,
  Michaelides, Morales, \& Markland}]{ceriotti2016nuclear}
Ceriotti, M., Fang, W., Kusalik, P.~G., {et~al.} 2016, Chem. Rev., 116, 7529,
  \dodoi{10.1021/acs.chemrev.5b00674}

\bibitem[{De~Simone {et~al.}(2022)De~Simone, Ceccarelli, Codella, Svoboda,
  Chandler, Bouvier, Yamamoto, Sakai, Yang, Caselli, {et~al.}}]{de2022tracking}
De~Simone, M., Ceccarelli, C., Codella, C., {et~al.} 2022, ApJL, 935, L14,
  \dodoi{https://doi.org/10.3847/2041-8213/ac85af}

\bibitem[{Dulieu {et~al.}(2010)Dulieu, Amiaud, Congiu, Fillion, Matar, Momeni,
  Pirronello, \& Lemaire}]{dulieu2010experimental}
Dulieu, F., Amiaud, L., Congiu, E., {et~al.} 2010, A\&A, 512, A30,
  \dodoi{10.1051/0004-6361/200912079}

\bibitem[{Enrique-Romero {et~al.}(2019)Enrique-Romero, Rimola, Ceccarelli,
  Ugliengo, Balucani, \& Skouteris}]{enrique2019reactivity}
Enrique-Romero, J., Rimola, A., Ceccarelli, C., {et~al.} 2019, ACS Earth Space
  Chem., 3, 2158, \dodoi{10.1021/acsearthspacechem.9b00156}

\bibitem[{Enrique-Romero {et~al.}(2022)Enrique-Romero, Rimola, Ceccarelli,
  Ugliengo, Balucani, \& Skouteris}]{enrique2022quantum}
---. 2022, ApJ Suppl. Ser., 259, 39,
  \dodoi{https://doi.org/10.3847/1538-4365/ac480e}

\bibitem[{Ferrero(2022)}]{sferrero_dissipation_NH3}
Ferrero, S. 2022, {Supporting Information for "Where does the energy go during
  the interstellar NH$_3$ formation on water ice? A computational study"}, 1.0,
   Zenodo, \dodoi{10.5281/zenodo.7115984}

\bibitem[{Fraser {et~al.}(2004)Fraser, Collings, Dever, \&
  McCoustra}]{fraser2004using}
Fraser, H.~J., Collings, M.~P., Dever, J.~W., \& McCoustra, M.~R. 2004, MNRAS,
  353, 59, \dodoi{10.1111/j.1365-2966.2004.08038.x}

\bibitem[{Fredon \& Cuppen(2018)}]{fredon2018molecular}
Fredon, A., \& Cuppen, H. 2018, PCCP, 20, 5569,
  \dodoi{https://doi.org/10.1039/C7CP06136F}

\bibitem[{Fredon {et~al.}(2017)Fredon, Lamberts, \& Cuppen}]{fredon2017energy}
Fredon, A., Lamberts, T., \& Cuppen, H. 2017, ApJ, 849, 125,
  \dodoi{https://doi.org/10.3847/1538-4357/aa8c05}

\bibitem[{Fuchs {et~al.}(2009)Fuchs, Cuppen, Ioppolo, Romanzin, Bisschop,
  Andersson, Van~Dishoeck, \& Linnartz}]{fuchs2009hydrogenation}
Fuchs, G., Cuppen, H., Ioppolo, S., {et~al.} 2009, A\&A, 505, 629,
  \dodoi{10.1051/0004-6361/200810784}

\bibitem[{Garrod {et~al.}(2007)Garrod, Wakelam, \& Herbst}]{garrod2007non}
Garrod, R., Wakelam, V., \& Herbst, E. 2007, A\&A, 467, 1103,
  \dodoi{10.1051/0004-6361:20066704}

\bibitem[{Goedecker {et~al.}(1996)Goedecker, Teter, \&
  Hutter}]{goedecker1996separable}
Goedecker, S., Teter, M., \& Hutter, J. 1996, Phys. Rev. B, 54, 1703,
  \dodoi{https://doi.org/10.1103/PhysRevB.54.1703}

\bibitem[{Grimme {et~al.}(2010)Grimme, Antony, Ehrlich, \&
  Krieg}]{grimme2010consistent}
Grimme, S., Antony, J., Ehrlich, S., \& Krieg, H. 2010, J. Chem. Phys., 132,
  154104, \dodoi{http://dx.doi.org/10.1063/1.3382344}

\bibitem[{Grimme {et~al.}(2011)Grimme, Ehrlich, \& Goerigk}]{grimme2011effect}
Grimme, S., Ehrlich, S., \& Goerigk, L. 2011, J. Comp. Chem., 32, 1456,
  \dodoi{https://doi.org/10.1002/jcc.21759}

\bibitem[{Guti{\'e}rrez-Quintanilla {et~al.}(2021)Guti{\'e}rrez-Quintanilla,
  Layssac, Butscher, Henkel, Tsegaw, Grote, Sander, Borget, Chiavassa, \&
  Duvernay}]{gutierrez2021icom}
Guti{\'e}rrez-Quintanilla, A., Layssac, Y., Butscher, T., {et~al.} 2021, MNRAS,
  506, 3734, \dodoi{https://doi.org/10.1093/mnras/stab1850}

\bibitem[{Habershon \& Manolopoulos(2009)}]{habershon2009zero}
Habershon, S., \& Manolopoulos, D.~E. 2009, J. Chem. Phys., 131, 244518,
  \dodoi{https://doi.org/10.1063/1.3276109}

\bibitem[{Hama {et~al.}(2012)Hama, Kuwahata, Watanabe, Kouchi, Kimura, Chigai,
  \& Pirronello}]{hama2012mechanism}
Hama, T., Kuwahata, K., Watanabe, N., {et~al.} 2012, ApJ, 757, 185,
  \dodoi{10.1088/0004-637X/757/2/185}

\bibitem[{Hama \& Watanabe(2013)}]{hama2013surface}
Hama, T., \& Watanabe, N. 2013, Chem. Rev., 113, 8783,
  \dodoi{dx.doi.org/10.1021/cr4000978}

\bibitem[{Hattig {et~al.}(2012)Hattig, Klopper, Kohn, \&
  Tew}]{hattig2012explicitly}
Hattig, C., Klopper, W., Kohn, A., \& Tew, D.~P. 2012, Chem. Rev., 112, 4,
  \dodoi{https://doi.org/10.1021/cr200168z}

\bibitem[{Henning(2010)}]{henning2010cosmic}
Henning, T. 2010, ARA\&A, 48, 0, \dodoi{10.1146/annurev-astro-081309-130815}

\bibitem[{Hocuk \& Cazaux(2015)}]{hocuk2015interplay}
Hocuk, S., \& Cazaux, S. 2015, A\&A, 576, A49,
  \dodoi{10.1051/0004-6361/201424503}

\bibitem[{Huppert {et~al.}(2022)Huppert, Pl{\'e}, Bonella, Depondt, \&
  Finocchi}]{huppert2022simulation}
Huppert, S., Pl{\'e}, T., Bonella, S., Depondt, P., \& Finocchi, F. 2022, Appl.
  Sci., 12, 4756, \dodoi{https://doi.org/10.3390/app12094756}

\bibitem[{Jones {et~al.}(2013)Jones, Fanciullo, K{\"o}hler, Verstraete,
  Guillet, Bocchio, \& Ysard}]{jones2013evolution}
Jones, A., Fanciullo, L., K{\"o}hler, M., {et~al.} 2013, A\&A, 558, A62,
  \dodoi{10.1051/0004-6361/201321686}

\bibitem[{Jones {et~al.}(2017)Jones, K{\"o}hler, Ysard, Bocchio, \&
  Verstraete}]{jones2017global}
Jones, A., K{\"o}hler, M., Ysard, N., Bocchio, M., \& Verstraete, L. 2017,
  A\&A, 602, A46, \dodoi{10.1051/0004-6361/201630225}

\bibitem[{Jorgensen {et~al.}(1983)Jorgensen, Chandrasekhar, Madura, Impey, \&
  Klein}]{jorgensen1983comparison}
Jorgensen, W.~L., Chandrasekhar, J., Madura, J.~D., Impey, R.~W., \& Klein,
  M.~L. 1983, J. Chem. Phys., 79, 926, \dodoi{https://doi.org/10.1063/1.445869}

\bibitem[{K{\"u}hne {et~al.}(2020)K{\"u}hne, Iannuzzi, Del~Ben, Rybkin,
  Seewald, Stein, Laino, Khaliullin, Sch{\"u}tt, Schiffmann,
  {et~al.}}]{kuhne2020cp2k}
K{\"u}hne, T.~D., Iannuzzi, M., Del~Ben, M., {et~al.} 2020, J. Chem. Phys.,
  152, 194103, \dodoi{https://doi.org/10.1063/5.0007045}

\bibitem[{Lippert {et~al.}(1997)Lippert, Hutter, \&
  Parrinello}]{lippert1997hybrid}
Lippert, G., Hutter, J., \& Parrinello, M. 1997, Mol. Phys., 92, 477,
  \dodoi{https://doi.org/10.1080/002689797170220}

\bibitem[{Mangaud {et~al.}(2019)Mangaud, Huppert, Pl{\'e}, Depondt, Bonella, \&
  Finocchi}]{mangaud2019fluctuation}
Mangaud, E., Huppert, S., Pl{\'e}, T., {et~al.} 2019, JCTC, 15, 2863,
  \dodoi{10.1021/acs.jctc.8b01164}

\bibitem[{Markland \& Ceriotti(2018)}]{markland2018nuclear}
Markland, T.~E., \& Ceriotti, M. 2018, Nat. Rev. Chem., 2, 1,
  \dodoi{10.1021/acs.chemrev.5b00674}

\bibitem[{Melani {et~al.}(2021)Melani, Nagata, \&
  Saalfrank}]{melani2021vibrational}
Melani, G., Nagata, Y., \& Saalfrank, P. 2021, PCCP, 23, 7714,
  \dodoi{10.1039/d0cp03777j}

\bibitem[{Meyer \& Reuter(2014)}]{meyer2014modeling}
Meyer, J., \& Reuter, K. 2014, Angew. Chem. Int. Ed., 53, 4721,
  \dodoi{10.1002/anie.201400066}

\bibitem[{Minissale {et~al.}(2016{\natexlab{a}})Minissale, Dulieu, Cazaux, \&
  Hocuk}]{minissale2016dust}
Minissale, M., Dulieu, F., Cazaux, S., \& Hocuk, S. 2016{\natexlab{a}}, A\&A,
  585, A24, \dodoi{10.1051/0004-6361/201525981}

\bibitem[{Minissale {et~al.}(2016{\natexlab{b}})Minissale, Moudens, Baouche,
  Chaabouni, \& Dulieu}]{minissale2016hydrogenation}
Minissale, M., Moudens, A., Baouche, S., Chaabouni, H., \& Dulieu, F.
  2016{\natexlab{b}}, MNRAS, 458, 2953, \dodoi{10.1093/mnras/stw373}

\bibitem[{Mukherjee \& Barbatti(2022)}]{mukherjee2022hessian}
Mukherjee, S., \& Barbatti, M. 2022, JCTC, 18, 4109–4116,
  \dodoi{https://doi.org/10.1021/acs.jctc.2c00216}

\bibitem[{Najibi \& Goerigk(2018)}]{najibi2018nonlocal}
Najibi, A., \& Goerigk, L. 2018, JCTC, 14, 5725,
  \dodoi{https://doi.org/10.1021/acs.jctc.8b00842}

\bibitem[{Neese(2004)}]{neese2004definition}
Neese, F. 2004, J. Phys. Chem Solids, 65, 781,
  \dodoi{https://doi.org/10.1016/j.jpcs.2003.11.015}

\bibitem[{Neese {et~al.}(2020)Neese, Wennmohs, Becker, \&
  Riplinger}]{neese2020orca}
Neese, F., Wennmohs, F., Becker, U., \& Riplinger, C. 2020, J. Chem. Phys.,
  152, 224108, \dodoi{10.1063/5.0004608}

\bibitem[{Pantaleone {et~al.}(2021)Pantaleone, Enrique-Romero, Ceccarelli,
  Ferrero, Balucani, Rimola, \& Ugliengo}]{pantaleone2021h2}
Pantaleone, S., Enrique-Romero, J., Ceccarelli, C., {et~al.} 2021, ApJ, 917,
  49, \dodoi{10.3847/1538-4357/ac0142}

\bibitem[{Pantaleone {et~al.}(2020)Pantaleone, Enrique-Romero, Ceccarelli,
  Ugliengo, Balucani, \& Rimola}]{pantaleone2020chemical}
---. 2020, ApJ, 897, 56, \dodoi{10.3847/1538-4357/ab8a4b}

\bibitem[{Perdew {et~al.}(1996)Perdew, Burke, \&
  Ernzerhof}]{perdew1996generalized}
Perdew, J.~P., Burke, K., \& Ernzerhof, M. 1996, PRL, 77, 3865,
  \dodoi{https://doi.org/10.1103/PhysRevLett.77.3865}

\bibitem[{Perrero {et~al.}(2022)Perrero, Enrique-Romero, Mart{\'\i}nez-Bachs,
  Ceccarelli, Balucani, Ugliengo, \& Rimola}]{perrero2022non}
Perrero, J., Enrique-Romero, J., Mart{\'\i}nez-Bachs, B., {et~al.} 2022, ACS
  Earth Space Chem., 6, 496, \dodoi{10.1021/acsearthspacechem.1c00369}

\bibitem[{Peterson {et~al.}(2008)Peterson, Adler, \&
  Werner}]{Peterson_CCSDT_basis}
Peterson, K.~A., Adler, T.~B., \& Werner, H.-J. 2008, J. Chem. Phys., 128,
  084102, \dodoi{10.1063/1.2831537}

\bibitem[{Pezzella \& Meuwly(2019)}]{pezzella20192}
Pezzella, M., \& Meuwly, M. 2019, PCCP, 21, 6247, \dodoi{10.1039/c8cp07474g}

\bibitem[{Potapov {et~al.}(2020)Potapov, J\"ager, \& Henning}]{potapov2020ice}
Potapov, A., J\"ager, C., \& Henning, T. 2020, PRL, 124, 221103,
  \dodoi{10.1103/PhysRevLett.124.221103}

\bibitem[{Qasim {et~al.}(2018)Qasim, Chuang, Fedoseev, Ioppolo, Boogert, \&
  Linnartz}]{qasim2018formation}
Qasim, D., Chuang, K.-J., Fedoseev, G., {et~al.} 2018, A\&A, 612, A83,
  \dodoi{https://doi.org/10.1051/0004-6361/201732355}

\bibitem[{Rimola {et~al.}(2014)Rimola, Taquet, Ugliengo, Balucani, \&
  Ceccarelli}]{rimola2014combined}
Rimola, A., Taquet, V., Ugliengo, P., Balucani, N., \& Ceccarelli, C. 2014,
  A\&A, 572, A70, \dodoi{10.1051/0004-6361/201424046}

\bibitem[{Rittmeyer {et~al.}(2018)Rittmeyer, Bukas, \&
  Reuter}]{rittmeyer2018energy}
Rittmeyer, S.~P., Bukas, V.~J., \& Reuter, K. 2018, Adv. Phys. X, 3, 1381574,
  \dodoi{https://doi.org/10.1080/23746149.2017.1381574}

\bibitem[{Ruaud {et~al.}(2015)Ruaud, Loison, Hickson, Gratier, Hersant, \&
  Wakelam}]{ruaud2015modelling}
Ruaud, M., Loison, J., Hickson, K., {et~al.} 2015, MNRAS, 447, 4004,
  \dodoi{10.1093/mnras/stu2709}

\bibitem[{Shakouri {et~al.}(2018)Shakouri, Behler, Meyer, \&
  Kroes}]{shakouri2018analysis}
Shakouri, K., Behler, J., Meyer, J., \& Kroes, G.-J. 2018, J. Phys. Chem. C,
  122, 23470, \dodoi{10.1021/acs.jpcc.8b06729}

\bibitem[{Takahashi {et~al.}(1999)Takahashi, Masuda, \&
  Nagaoka}]{takahashi1999product}
Takahashi, J., Masuda, K., \& Nagaoka, M. 1999, ApJ, 520, 724,
  \dodoi{https://iopscience.iop.org/article/10.1086/307461/meta}

\bibitem[{Tinacci {et~al.}(2022)Tinacci, Germain, Pantaleone, Ferrero,
  Ceccarelli, \& Ugliengo}]{tinacci2022theoretical}
Tinacci, L., Germain, A., Pantaleone, S., {et~al.} 2022, ACS Earth Space Chem.,
  \dodoi{https://doi.org/10.1021/acsearthspacechem.2c00040}

\bibitem[{Upadhyay \& Meuwly(2021)}]{upadhyay2021energy}
Upadhyay, M., \& Meuwly, M. 2021, Front. Chem., 9,
  \dodoi{10.3389/fchem.2021.827085}

\bibitem[{Valeev(2004)}]{valeev2004improving}
Valeev, E.~F. 2004, Chem. Phys. Lett., 395, 190,
  \dodoi{10.1016/j.cplett.2004.07.061}

\bibitem[{Van~Dishoeck {et~al.}(2013)Van~Dishoeck, Herbst, \&
  Neufeld}]{van2013interstellar}
Van~Dishoeck, E.~F., Herbst, E., \& Neufeld, D.~A. 2013, Chem. Rev., 113, 9043,
  \dodoi{10.1021/cr4003177}

\bibitem[{VandeVondele \& Hutter(2007)}]{vandevondele2007gaussian}
VandeVondele, J., \& Hutter, J. 2007, J. Chem. Phys., 127, 114105,
  \dodoi{10.1063/1.2770708}

\bibitem[{Vasyunin \& Herbst(2013)}]{vasyunin2013reactive}
Vasyunin, A., \& Herbst, E. 2013, ApJ, 769, 34,
  \dodoi{10.1088/0004-637X/769/1/34}

\bibitem[{Vidali(2013)}]{vidali2013h2}
Vidali, G. 2013, Chem. Rev., 113, 8762, \dodoi{10.1021/cr400156b}

\bibitem[{Wakelam {et~al.}(2017)Wakelam, Bron, Cazaux, Dulieu, Gry, Guillard,
  Habart, Hornekaer, Morisset, Nyman, {et~al.}}]{wakelam2017h2}
Wakelam, V., Bron, E., Cazaux, S., {et~al.} 2017, Mol. Astrophys., 9, 1,
  \dodoi{10.1016/j.molap.2017.11.001}

\bibitem[{Watanabe {et~al.}(2010)Watanabe, Kimura, Kouchi, Chigai, Hama, \&
  Pirronello}]{watanabe2010direct}
Watanabe, N., Kimura, Y., Kouchi, A., {et~al.} 2010, ApJL, 714, L233,
  \dodoi{10.1088/2041-8205/714/2/L233}

\bibitem[{Watanabe \& Kouchi(2002)}]{watanabe2002efficient}
Watanabe, N., \& Kouchi, A. 2002, ApJ, 571, L173, \dodoi{10.1086/341412}

\bibitem[{Watanabe \& Kouchi(2008)}]{watanabe2008ice}
---. 2008, Prog. Surf. Sci., 83, 439, \dodoi{10.1016/j.progsurf.2008.10.001}

\bibitem[{Watanabe {et~al.}(2004)Watanabe, Nagaoka, Shiraki, \&
  Kouchi}]{watanabe2004hydrogenation}
Watanabe, N., Nagaoka, A., Shiraki, T., \& Kouchi, A. 2004, ApJ, 616, 638,
  \dodoi{10.1086/424815}

\bibitem[{Weigend \& Ahlrichs(2005)}]{weigend2005balanced}
Weigend, F., \& Ahlrichs, R. 2005, PCCP, 7, 3297, \dodoi{10.1039/B508541A}

\bibitem[{Zamirri {et~al.}(2019)Zamirri, Ugliengo, Ceccarelli, \&
  Rimola}]{zamirri2019quantum}
Zamirri, L., Ugliengo, P., Ceccarelli, C., \& Rimola, A. 2019, ACS Earth Space
  Chem., 3, 1499, \dodoi{https://doi.org/10.1021/acsearthspacechem.9b00082}

\end{thebibliography}
\bibliographystyle{aasjournal}



\end{document}